\documentclass[reprint,amsmath,amssymb,aps,floatfix]{revtex4-2}

\usepackage{graphicx}
\usepackage{dcolumn}
\usepackage{bm}
\usepackage[dvipsnames]{xcolor}

\usepackage{tikz}
\usetikzlibrary{positioning, arrows.meta}
\usepackage{ragged2e}
\usepackage{multirow}
\usepackage{placeins}
\usepackage{enumitem}
\usepackage{relsize}
\usepackage{overpic}
\usepackage[export]{adjustbox}

\usepackage{mathtools}
\usepackage{makecell}
\usepackage{tabularx}
\usepackage{booktabs}
\usepackage{array}
\usepackage{longtable}

\usepackage[linesnumbered,ruled,vlined]{algorithm2e}

\usepackage{comment}
\usepackage{textcase}
\usepackage[normalem]{ulem}

\usepackage{hyperref}

\usepackage{amsthm}

\theoremstyle{remark}

\begin{document}

\preprint{APS/123-QED}

\title{Leveraging Population Dynamics to Steer Efficient Search in Large-Scale Combinatorial Optimization}

\author{Nikhat Khan, Ridge Redding, Nikhil Shukla}
\affiliation{%
University of Virginia, Charlottesville, VA 22903, USA
}%

\begin{abstract}

    Combinatorial optimization problems pose substantial computational challenges because their feasible solution spaces grow exponentially with problem size. This paper presents a GPU-accelerated augmented Population Annealing Monte Carlo (PAMC) framework for large-scale graph-partitioning problems, with emphasis on Max-Cut and Max-$K$-Cut. The proposed framework extends conventional PAMC by coupling population-based resampling with two stagnation-driven mechanisms: adaptive temperature control and energy-preserving nonlocal cluster moves. By using population-level optimization history as feedback, these mechanisms regulate the balance between exploration and refinement by reheating stalled populations and enabling collective transitions across locally confined regions of the solution space. Experiments on G-set benchmark instances show that the augmented PAMC framework achieves competitive or lower time-to-solution than reported state-of-the-art baselines on several large Max-Cut instances, while matching or improving solution quality under comparable runtime budgets. The solver also discovers a new best-known solution for the G63 Max-Cut instance and scales to a fully connected $100{,}000$-spin Ising instance. For Max-$3$-Cut, the same framework establishes new best-known solutions on 36 G-set instances, demonstrating its applicability beyond binary Ising formulations. These results indicate that feedback-controlled population dynamics provide an effective and scalable strategy for steering stochastic search in large-scale combinatorial optimization.

\end{abstract}
                         
\maketitle

\section{Introduction}

Combinatorial optimization problems (COPs) are central to many applications across industry and research, including machine learning, operations research, materials design, and scientific computing. Many COPs can be reformulated as Ising-energy minimization problems. A standard pairwise Ising Hamiltonian can be written as $H(s)=-\sum{J_{ij}s_is_j}$,
where $\mathbf{s} = (s_1,\dots,s_N)$ denotes a spin configuration, each spin variable $s_i \in \{-1,+1\}$
represents one of two possible spin states, $J_{ij}$ is the coupling coefficient between spins $i$ and $j$, and $N$ is the number of spins. After mapping the COP objective function to an Ising Hamiltonian, finding the ground state, or minimum-energy configuration of this Hamiltonian, corresponds to solving the original COP. However, many COPs are nondeterministic polynomial-time (NP)-hard, and the size of the feasible solution space grows exponentially with problem size, creating a combinatorial explosion that makes even moderately sized instances computationally challenging for classical digital solvers in terms of runtime and memory requirements.

In this work, we focus on Max-Cut, one of the most widely used benchmark problems for comparing combinatorial optimization solvers, and its multi-state generalization, Max-$K$-Cut, which extends this benchmark beyond binary decision variables \cite{seo2015edgeset}\cite{benlic2011multilevel}. Given a graph, Max-Cut seeks a partition of the vertex set into two disjoint subsets such that the weight of the edges crossing between the subsets is maximized. Its objective function admits a direct mapping to an Ising Hamiltonian, making it a natural and convenient testbed for Ising-based optimization frameworks. Max-$K$-Cut extends this objective by partitioning the vertices into $K>2$ subsets and maximizing the total weight of edges with endpoints in different subsets. This multi-state partitioning structure makes Max-$K$-Cut relevant to a broader class of partitioning and assignment problems that arise in applications such as clustering, frequency resource allocation, and scheduling. However, unlike Max-Cut, Max-$K$-Cut is not natively an Ising problem, and embedding it in an Ising form requires an additional mapping step that introduces auxiliary nodes and edges. This embedding inflates the problem size. 

Among the many algorithmic paradigms applied to COPs, Markov-chain Monte Carlo (MCMC) methods~\cite{Metropolis1953equation} are especially relevant to energy-based optimization. Rather than exhaustively searching an exponentially large configuration space, MCMC generates stochastic trajectories whose stationary distribution is chosen to favor desirable configurations. In optimization settings, this target distribution is commonly chosen to be the Boltzmann distribution, so that lower-energy configurations are sampled with higher probability at a given temperature.
A prominent realization of this principle is simulated annealing (SA) \cite{Kirkpatrick1983optimization}, in which a single Markov chain is evolved under a gradually decreasing temperature schedule, progressively biasing the chain toward low-energy states. However, since the search is carried out along a single trajectory, SA can struggle to move between distant basins in rugged energy landscapes. Although independent SA runs are easily parallelized, they do not exchange information, making their performance sensitive to initial conditions, annealing schedules, and other solver parameters. 

Two parallel descendants of SA address this limitation by replacing the independent-runs structure with a coordinated ensemble of replicas that can share information~\cite{Iba2001population}. The first one is parallel tempering (PT), which runs multiple replicas of the same system at different temperatures, so high-temperature replicas explore broadly, while low-temperature replicas concentrate on low-energy states~\cite{Earl2005parallel}. Periodic swaps between replicas allow configurations to move across the temperature ladder, helping low-temperature chains escape local minima and explore rugged landscapes more effectively. The second one is Population Annealing Monte Carlo (PAMC), the focus of this paper, which follows a different ensemble-based strategy \cite{Hukushima2003population}. It evolves a population of replicas through a sequence of increasing inverse temperature and performs an energy-dependent resampling step at each temperature. This resampling preferentially replicates low-energy configurations while removing higher-energy ones, allowing the population to approximate a Gibbs distribution at each stage. 

\begin{figure*}[t]
    \centering
    \includegraphics[width=1\linewidth]{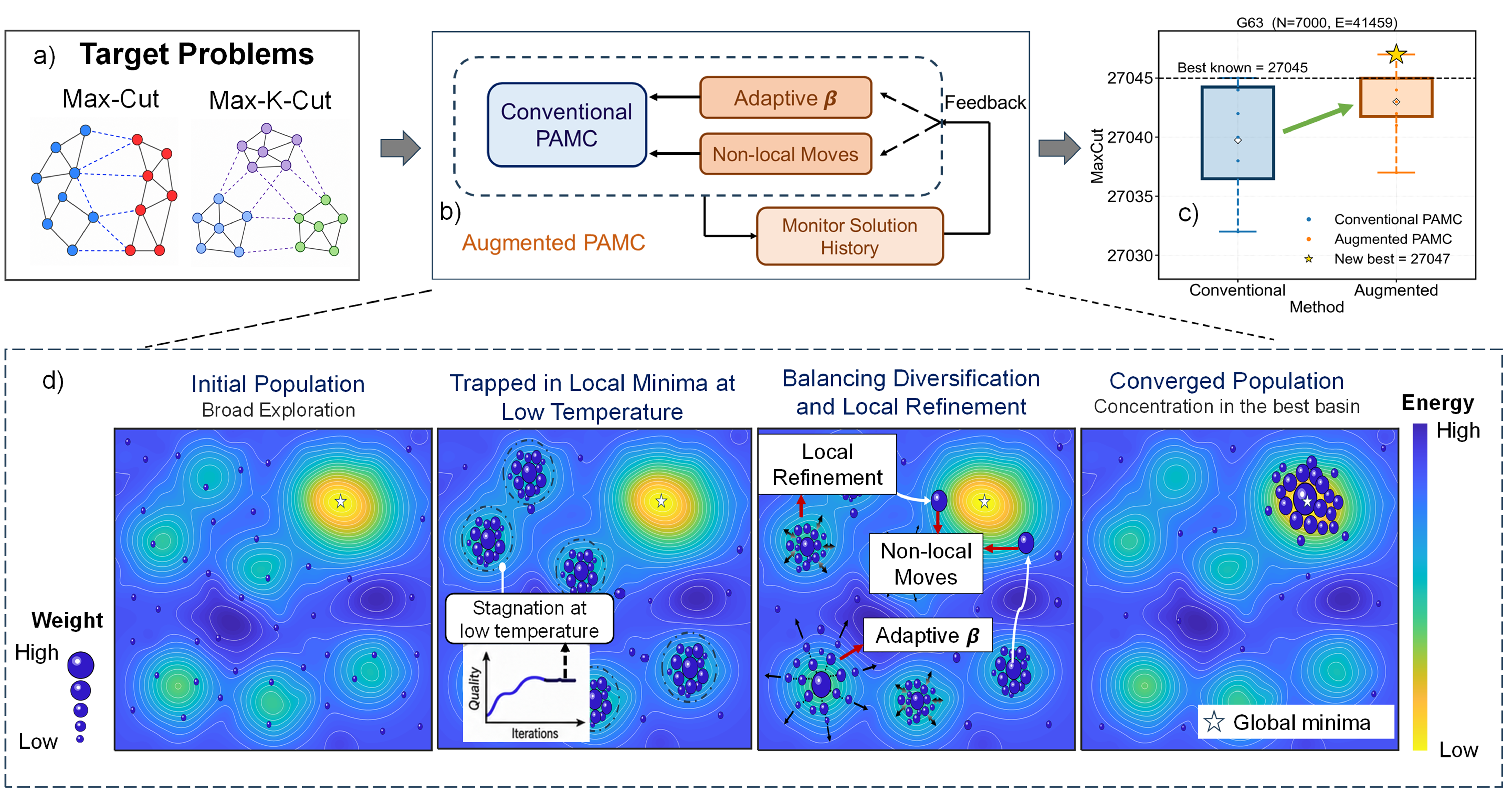}
    \caption{\justifying Overview of the augmented PAMC framework proposed here. (a) The framework targets graph partitioning COPs, including Max-Cut and Max-$K$-Cut. (b) An augmented PAMC framework with a feedback-driven control loop that monitors solution history and activates adaptive temperature control and non-local cluster moves when search progress stalls. (c) Representative Max-Cut result on the G63 instance, illustrating improved solution quality obtained by the augmented PAMC relative to conventional PAMC, including discovery of a new best-known cut value for G63. (d) Conceptual illustration of the population dynamics on an energy landscape. During annealing, replicas initially sample diverse regions of the landscape but may concentrate in local minima. Adaptive reheating and nonlocal cluster moves help restore exploration and guide the population toward high-quality solution regions. }
    \label{fig:common_framework}
\end{figure*}

Although PT has been extensively exploited as a COP solver \cite{Huang2024ising, zhu2020borealis, delacourKerem2025two, nikhar2024allKerem}, PAMC remains less explored in optimization-focused settings despite its natural suitability for accelerator-based hardware. Prior work on PAMC as a COP solver includes \cite{Wang2015comparing}, which showed that both PT and PAMC outperform SA on hard spin-glass optimization benchmarks, with PAMC achieving performance comparable to PT. Since then, PAMC has been improved along several directions, including adaptive temperature schedules based on culling fraction and energy variance \cite{Amey2018analysis}, energy-histogram overlap \cite{Barash2017gpu}, or optimal annealing schedules \cite{barzegar2024optimal}; massively parallel GPU and MPI implementations \cite{Barash2017gpu, barzegar2018optimization}; weighted averaging \cite{ebert2022weighted} and error diagnostics \cite{weigel2021understanding}; and more recently, topological-defect-driven nonlocal updates \cite{Cirauqui2024population}. However, many of these advances have been demonstrated primarily on spin-glass or gauge-model benchmarks, motivating a closer evaluation of PAMC for graph-partitioning COPs such as Max-Cut and Max-$K$-Cut.
 
To address this question, we introduce an augmented PAMC framework for Max-Cut and Max-$K$-Cut that coordinates adaptive temperature control and energy-preserving nonlocal cluster moves through a single stagnation-driven controller. Fig.~\ref{fig:common_framework} presents an overview of the proposed augmented PAMC framework. The controller tracks the population-wide best-so-far cut values over a finite memory window and uses lack of improvement as a signal of stagnation, allowing the population to increase exploration when progress stalls. The contribution of this work lies in using this common stagnation-detection signal to coordinate both adaptive reheating and nonlocal moves within a unified PAMC framework.
Adaptive temperature control has been studied in SA \cite{Ingber1989VFSR, Karabin2020adaptive}, PT \cite{Katzgraber2006feedback, Miasojedow2013adaptive}, and PAMC \cite{Barash2017gpu, Amey2018analysis}. Prior PT and PAMC schemes typically rely on sampling-equilibration metrics to tune the temperature ladder or adjust the inverse-temperature step size, whereas reheating mechanisms based on optimization progress have, by contrast, mainly appeared in SA \cite{boese1994best}. Energy-preserving nonlocal cluster moves have likewise been studied in several forms. Houdayer's isoenergetic cluster move \cite{Houdayer2001cluster} constructs clusters from disagreement domains between two replicas in two-dimensional spin glasses, while \cite{Zhu2015isoenergetic} generalized this idea to spin glasses in any space dimension. More recently, \cite{Cirauqui2024population} introduced topological defect-driven nonlocal cluster updates within PA for the three-dimensional model, derived from a random 3D Ising gauge theory in plaquette representation. Our nonlocal cluster move differs from these prior approaches in how it is constructed and applied. 

Furthermore, relatively few solvers address Max-Cut and Max-$K$-Cut within a single framework. Several of the strongest classical software heuristics for Max-Cut—including Breakout Local Search (BLS) \cite{Benlic2013BLS}, Global Equilibrium Search with path relinking (GES-PR) \cite{Shylo2015teams}, and the Large Population Island (LPI) framework \cite{Goudet2024island} —are focused primarily on the ($K$=2) case. For Max-$K$-Cut, the Multiple-Operator Heuristic (MOH) \cite{Ma2017MOH} remains one of the strongest classical heuristics for general $K$ and provides an important benchmark for solution quality. In addition, Parameterized Local Search (PLS) has reported competitive Max-$K$-Cut results on benchmark instances, using both local-search based and ILP variants, making it a relevant classical baseline for evaluating solution quality \cite{garvardt2024parameterized}. Notably, the reported MOH and PLS results are based on CPU implementations. On the other hand, GPU-based dynamical Ising solvers such as Toshiba's Simulated Bifurcation Machine (SBM) \cite{Goto2021highperformance} achieve competitive time-to-solution for large Ising problems and have been demonstrated at scales up to one million variables. However, SBM has primarily focused on binary optimization problems. In addition, specialized solvers such as Cosm have demonstrated strong performance on sparse Ising optimization instances~\cite{zick2026cosm}. In parallel, p-bit-based probabilistic computing has emerged as a promising approach for Ising optimization, since many existing platforms are built from binary stochastic units \cite{Camsari2017pbits, Aadit2022massively, Borders2019factorization, yang2025250}. Recent work has extended this paradigm to multi-state problems by designing $K$-state p-bit engines~\cite{Bashar2024kstate, Cheong2026potts, Duffee2025pdits}. Complementary to these efforts, the GPU-accelerated augmented PAMC framework presented here provides a scalable algorithmic approach that supports both Max-Cut and Max-$K$-Cut within a common framework. 
The key contributions of this work are summarized as follows:

\begin{itemize}[leftmargin=*]
    \item  We augment conventional PAMC with a stagnation-driven temperature control and a uniquely implemented non-local cluster move strategy, yielding a common framework for solving both Max-Cut and Max-$K$-Cut.
    \item  We implement the framework on a
    GPU, exploiting the inherent parallelism of its replica population to accelerate
    the solver and achieve a lower time-to-solution than the state-of-the-art on several large G-set instances at matched solution quality.
    \item The solver establishes a new best-known
    Max-Cut solution for the G63 instance and new best-known Max-3-Cut solutions for 36 G-set instances.
\end{itemize}

\section{GPU Implementation and Benchmark Setup}
The augmented PAMC algorithm is implemented on a graphics processing unit (GPU) using CUDA, leveraging optimized libraries such as cuRAND, cuDNN, cuBLAS, and Thrust. All GPU kernels are executed on the University of Virginia High Performance Computing (HPC) cluster \cite{uva_research_computing}. The G-set benchmarking experiments were performed on an NVIDIA RTX A6000 GPU (compute capability 8.6), while the large Ising problem evaluation was performed using an eight-GPU NVIDIA A100 (compute capability 8.0).

We benchmark the proposed framework on instances from the widely used G-set suite~\cite{Gset_HelmbergRendl_Stanford}, a standard testbed for large-scale graph-partitioning solvers. The benchmark contains instances G1--G81, with graph sizes ranging from $800$ to $20{,}000$ vertices and approximately $4{,}000$ to $40{,}000$ edges, including both unipolar $(+1)$ and bipolar $(\pm 1)$ edge weights. Because best-known solutions for several G-set instances continue to be refined, the suite remains an active and informative benchmark for evaluating new combinatorial-optimization solvers.

In all benchmarking experiments, including both Max-Cut and Max-$3$-Cut, the algorithm was executed for each graph $G_i$ using the same replica count (=512) and inverse-temperature spacing \(\Delta\beta = 0.02\) across all graphs. The number of iterations was also fixed across all graphs, with $2000$ iterations for Max-Cut and $400$ iterations for Max-$3$-Cut, and the system was reset after every $400$ iterations.
 At the same time, the number of Monte Carlo sweeps for Max-Cut was scaled with the graph size, ranging from $10$ to $400$ sweeps, and was kept constant at $100$ for Max-3-Cut across all graphs.

\section{Results}

\subsection{Population Annealing Monte Carlo}
\begin{figure}[h]
    \centering
    \includegraphics[width=1\linewidth]{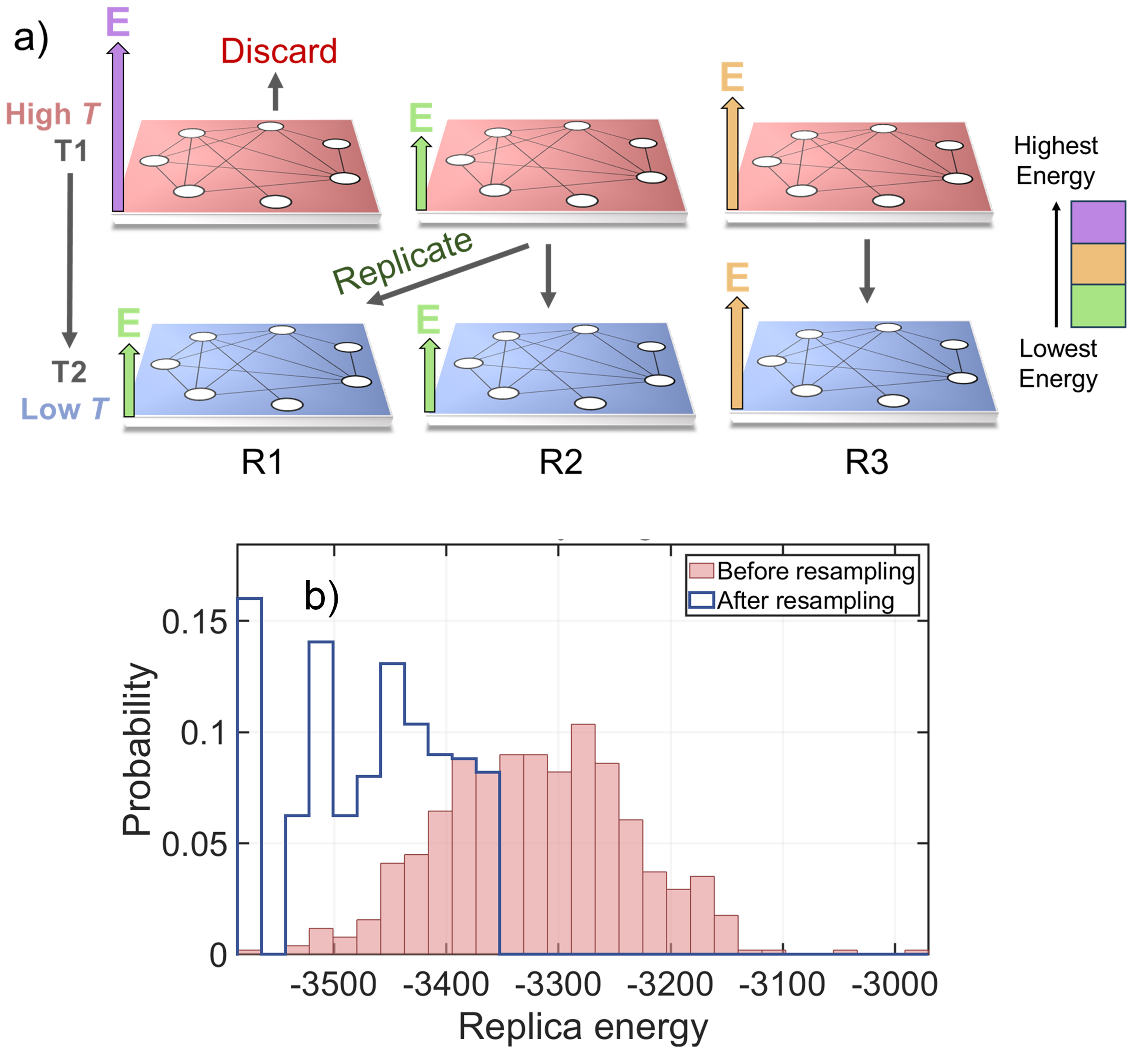}
    \caption{\justifying (a) Schematic illustration of the energy-dependent resampling step in PAMC. Each panel represents a replica of the system, with energy $E$ indicated by the vertical arrow and color shading from high energy (top, purple) to low energy (bottom, green). As the system is cooled from temperature T1 to T2, higher-energy replicas are more likely to be discarded, whereas lower-energy replicas are replicated. (b) Replica-energy distributions before and after resampling, showing that resampling shifts the population toward lower energies.}
    \label{fig:resampling}
\end{figure}

PAMC, first introduced by Hukushima and Iba in 2003, is an advanced population-based extension of Monte Carlo. Rather than propagating a single Markov chain, PAMC evolves a population of spin configurations simultaneously along an annealing schedule. In its native formulation, a population of $R$ independent system replicas is randomly initialized at a high initial temperature $\beta$ and then gradually cooled. While this aspect of the algorithm is similar to simulated annealing of the $R$ replicas with different initial conditions, population annealing uniquely employs a periodic resampling step to maintain a Gibbs-like energy distribution for the ensemble as demonstrated in Fig. \ref{fig:resampling}. 

The reweighting factor for a replica $j$ with energy $E_j$ is $\exp(-\Delta\beta E_j)$, where $\Delta \beta = \beta_{i+1} - \beta_i$. The number of copies of replica $j$ is then expressed as
\begin{equation}
    \tau_j (\beta_i, \beta_{i+1}) = \frac{e^{-(\beta_{i+1} - \beta_i) E_j}}{Q(\beta_i, \beta_{i+1})},
\end{equation}
where $Q(\beta_i, \beta_{i+1})$ is the normalization factor that enforces $\sum_j \tau_j = 1$,
\begin{equation}
    Q(\beta_i, \beta_{i+1}) = \frac{1}{R} \sum_{j=1}^{R} e^{-(\beta_{i+1} - \beta_i) E_j}.
\end{equation}
During resampling, low-energy replicas are preferentially replicated, while high-energy replicas are more likely to be removed from the population.

After resampling, the replicas become highly correlated. To decorrelate them, each replica undergoes a fixed number of Monte Carlo sweeps using a stochastic local spin-update rule, equivalent to the Gibbs sampler. This update rule is mathematically equivalent to the stochastic activation used in p-bit neurons, where the output state fluctuates probabilistically according to its input bias, with the effective temperature controlling the strength of the stochastic response \cite{Camsari2017pbits}. Specifically, for a spin configuration $\mathbf{s}$ and site $k$, the local effective field is
\begin{equation}
    h_k = \sum_{j} J_{kj} s_j,
\end{equation}
and the spin is updated according to
\begin{equation}
    s_k \leftarrow \operatorname{sgn}\!\left[\tanh(\beta h_k) - (2r - 1)\right], \quad r \sim U(0,1).
\end{equation}
where $r$ is drawn from a uniform random distribution over the interval $[0, 1]$.
During each Monte Carlo sweep, all spins were sequentially updated once in a randomized order, ensuring unbiased sampling of the configuration space. 

Population annealing collectively utilizes the information from the samples in parallel to improve the effectiveness of the search.
By maintaining population diversity, PAMC achieves broader exploration of the solution landscape while still reinforcing promising candidates.\\

\subsection{PAMC and SA}
\begin{figure}[h]
    \centering
    \includegraphics[width=1\linewidth]{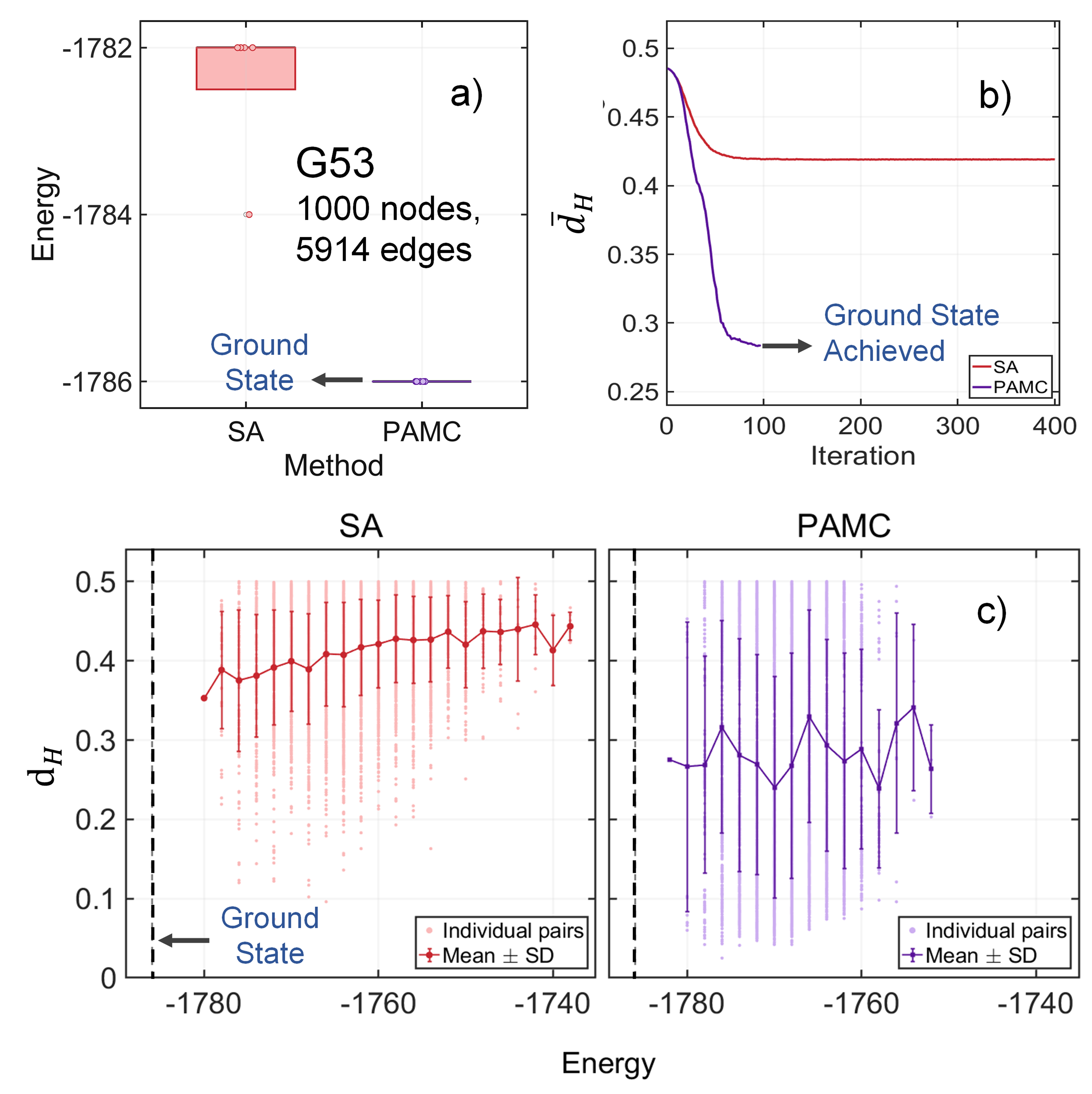}
    \caption{\justifying Comparison of PAMC and SA on the G53 (N: 1000; E: 5914) Max-Cut instance using replicas (=512). (a) Best Max-Cut value obtained by each method. (b) Evolution of $\bar{d}_{\mathrm H}$ among replicas across annealing iterations. (c) Distribution summary (mean and standard deviation) of the ${d}_{\mathrm H}$ among replicas with the same final energies.}
    \label{fig:discussion_fig_1}
\end{figure}

To elucidate the effect of resampling on the population dynamics, we compared Population Annealing Monte Carlo (PAMC) with Simulated Annealing (SA) baseline. In the SA baseline, $R$ replicas were initialized randomly and evolved independently in parallel. Both methods used the same Gibbs single-spin update rule and the same initialization protocol within each run, so that the algorithmic distinction was the presence of the resampling step in PAMC. Each method was evaluated over five independent runs, with different random seeds used across runs.

The results in Fig.~\ref{fig:discussion_fig_1} indicate that resampling significantly contributes to the improved performance of PAMC over SA on the G53 instance. As shown in Fig.~\ref{fig:discussion_fig_1}(a), PAMC attains a higher Max-Cut value than SA under otherwise matched update and initialization conditions. Since the local stochastic dynamics are held fixed between the two methods, the observed difference is consistent with the population-level selection mechanism introduced by resampling.

The corresponding evolution of the mean normalized pairwise Hamming distance ($\bar{d}_{\mathrm H}$), shown in Fig.~\ref{fig:discussion_fig_1}(b), further illustrates this effect, where $\bar{d}_{\mathrm H}$ is computed by averaging the normalized pairwise Hamming distances over all replica pairs. Details of the normalized pairwise Hamming distance (${d}_{\mathrm H}$) calculation are provided in Supplementary Material \ref{sec:methods_hamming_distance}. In the PAMC approach, $\bar{d}_{\mathrm H}$ decreases rapidly during the annealing process, indicating increasing structural concentration of the replica population. This concentration occurs concurrently with the attainment of the ground-state cut value, suggesting that the population is being selectively concentrated around high-quality regions of the configuration space rather than collapsing arbitrarily. In contrast, the SA replicas exhibit a larger $\bar{d}_{\mathrm H}$ and quickly approach a plateau, indicating that independently evolving trajectories remain more widely dispersed in configuration space and lack a mechanism to propagate favorable configurations across the ensemble.

Fig.~\ref{fig:discussion_fig_1}(c) provides a complementary view by examining the distribution of ${d}_{\mathrm H}$ among replicas with the same energies at the final iteration. For SA, replicas with identical final energy exhibit a larger mean ${d}_{\mathrm H}$, indicating that independent trajectories can terminate in structurally distinct configurations of comparable quality. In PAMC, the lower mean ${d}_{\mathrm H}$ at comparable energies indicates stronger population concentration within favorable regions. At the same time, the finite standard deviation shows that the population does not collapse to a single configuration, but retains structural variability within these regions. Taken together, Fig.~\ref{fig:discussion_fig_1}(b,c) suggests that resampling promotes population-level refinement, while subsequent MCMC sweeps preserve sufficient variability to maintain local exploration.

\subsection{Augmented PAMC}

\begin{figure*}[t]
    \centering

    \begin{minipage}[b]{0.49\textwidth}
        \centering
        \includegraphics[width=0.9\linewidth]{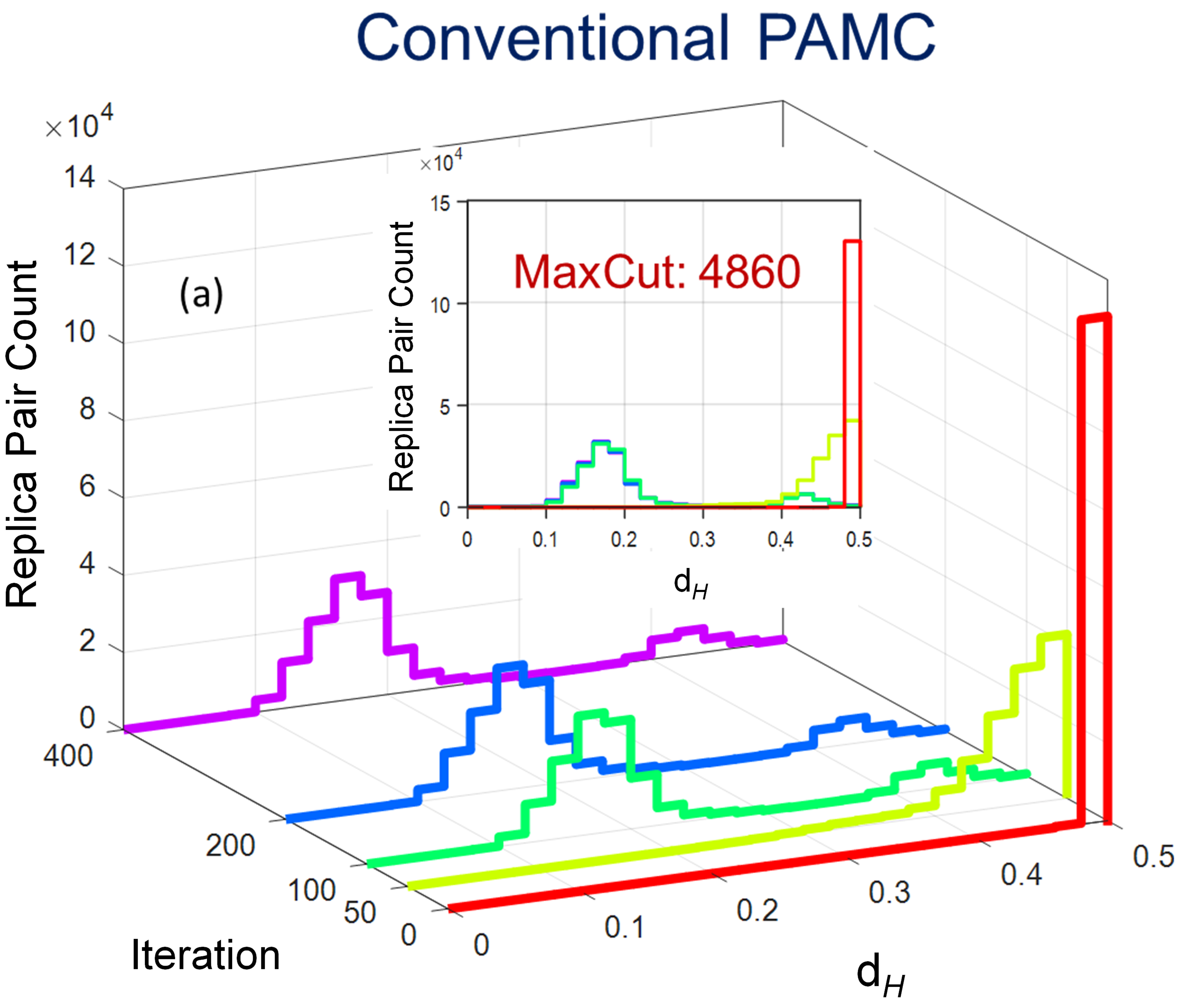}
    \end{minipage}
    \hfill
    \begin{minipage}[b]{0.49\textwidth}
        \centering
        \includegraphics[width=0.9\linewidth]{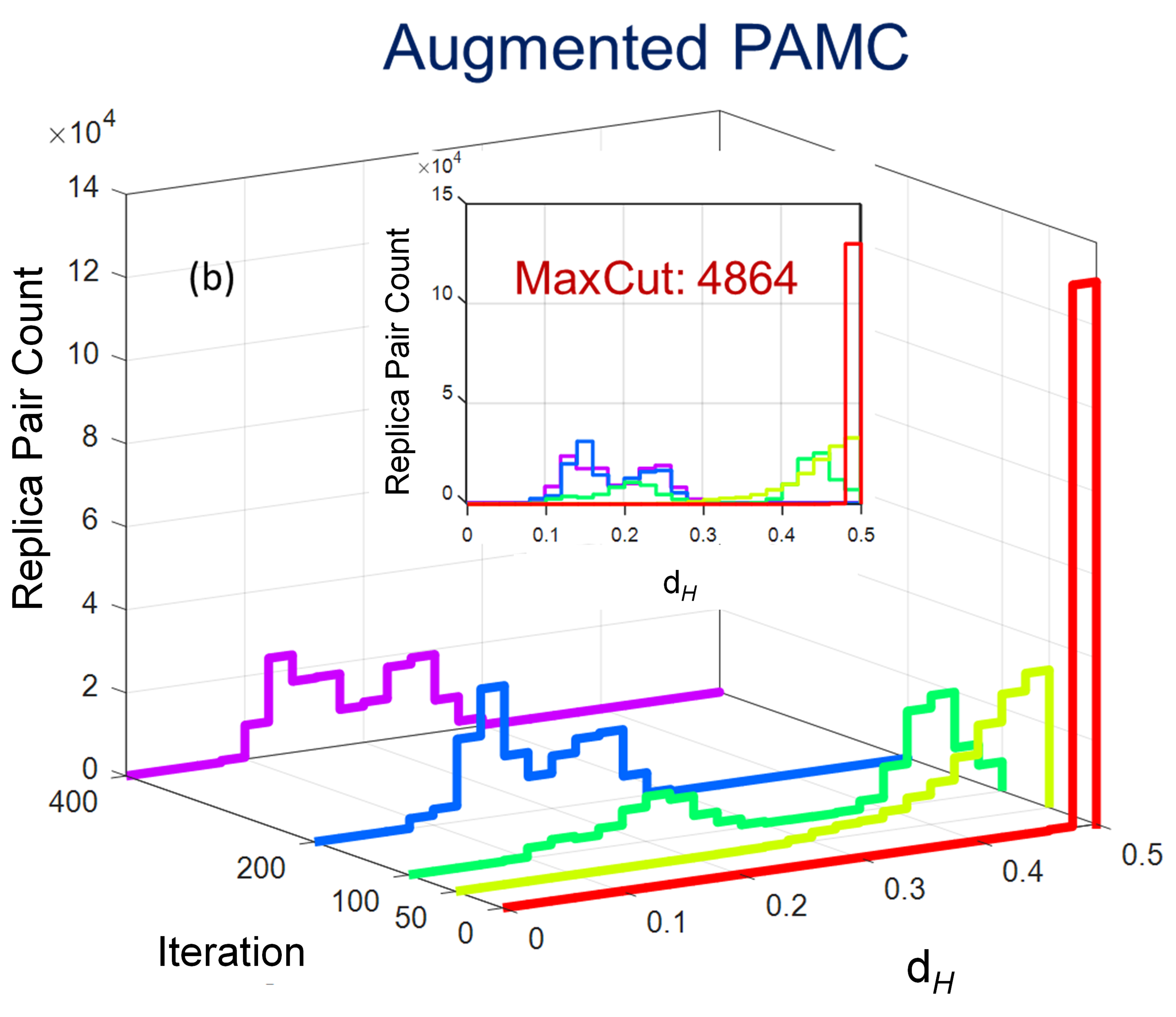}
    \end{minipage}

    \caption{Evolution of replica diversity during annealing for the G62 (N: 7000; E: 14000) instance using replicas (=512). The distributions show the normalized pairwise Hamming distances between all replica pairs at selected annealing iterations for (a) conventional PAMC and (b) augmented PAMC. The insets show frontal views of the same distributions to better visualize population spread. The augmented PAMC maintains a richer diversity over intermediate distances and attains a higher-quality solution (Max-Cut 4864) than the conventional PAMC (Max-Cut 4860). }
    \label{fig:g62_2x2}
\end{figure*}

\begin{algorithm}[t]
\DontPrintSemicolon
\SetKwInOut{Input}{Input}\SetKwInOut{Output}{Output}
\SetKwFunction{Cut}{Cut}\SetKwFunction{Diversity}{Diversity}
\caption{Augmented PAMC with Adaptive Temperature and Non-local Cluster Moves}
\label{pseudocode}
\Input{
Coupling matrix $J$; population size $R$; sweeps per step $S$;
target cut $\mathrm{MC}^\star$; reset threshold $z_{\rm rst}$;
initial inverse temperature $\beta_0$; annealing increment $\Delta\beta$;
cluster-move cooldown $\tau$; number of top configurations $n$
}
\Output{Best cut value $M^\ast$}

\textbf{Initialization:};
Generate a randomly initialized population $\mathcal{P}$ of $R$ replicas;\\
Set $z \gets 1$;\quad $\text{cycle}\gets 0$;\quad $\beta \gets \beta_0$;\quad $\Delta \gets \Delta\beta$\;
Initialize $M^\ast \gets 0$;\quad enable-cluster-move $\gets$ \textbf{false};\quad cooldown $\gets 0$\;

\While{$M^\ast < \mathrm{MC}^\star$}{

  \If{$z > 1$}{
    Resample $\mathcal{P}$ by Boltzmann weights of the current energies with increment $\Delta$\;
    \tcp{Non-local Cluster Moves}
    \If{\textnormal{enable-cluster-move} \textbf{and} \textnormal{cooldown} $=0$}{
      Apply non-local cluster moves to the $n$ top configurations\;
      enable-cluster-move $\gets$ \textbf{false};\quad cooldown $\gets \tau$\;
    }
  }

  Run $S$ MCMC updates on every replica at inverse temperature $\beta$\;
  $M^\ast \gets \max\!\big(M^\ast,\ \max \Cut(\mathcal{P})\big)$\;
  Record $M^\ast$ in the recent-progress history\;

  \tcp{Adaptive temperature control}
  \If{\textnormal{cooldown} $=0$ \textbf{and} history is long enough}{
    \uIf{\Diversity{history} \textnormal{is very low}}{
      $\beta \gets 0.8\,\beta$;\quad $\Delta \gets \Delta\beta$;\quad enable-cluster-move $\gets$ \textbf{true}
    }
    \uElseIf{$\beta > \beta_0$}{
      $\Delta \gets -\Delta\beta$ 
    }
    \Else{
      \textbf{break} 
    }
  }

  \eIf{$z \ge z_{\rm rst}$}{
    \tcp{Restart annealing schedule}
    $\text{cycle}\gets \text{cycle}+1$;\quad $z \gets 1$;\quad $\beta \gets \beta_0$\;
    $\Delta \gets \Delta\beta$;\quad enable-cluster-move $\gets$ \textbf{false};\quad cooldown $\gets 0$\;
  }{
    $z \gets z+1$;\quad $\beta \gets \beta + \Delta$\;
    \If{\textnormal{cooldown} $> 0$}{cooldown $\gets$ cooldown $-1$}
  }
}
\Return $M^\ast$\;
\end{algorithm}

As noted earlier, the PAMC framework was further extended with two enhancements as summarized in Algorithm~\ref{pseudocode} to improve the solution quality and convergence speed.  
\begin{itemize}[leftmargin=*]
    \item \textbf{Non-local Cluster Moves (NCM)}: A known limitation of local spin-flip updates in MCMC and conventional PAMC is that the system can get trapped in clusters of states separated by barriers that single-spin updates cannot efficiently overcome. To mitigate this, the non-local move strategy introduces non-local updates by identifying iso-sites, that is, variables (spins) where the local field is exactly balanced, so flipping them does not alter the system energy. From these iso-sites, a maximal independent set (MIS) is constructed using randomized selection, in which the iso-sites are randomly permuted, the first site is selected, and each subsequent site is added only if it has zero coupling to all previously selected sites. This procedure is repeated for a fixed number of times, and the largest resulting subset of independent spins is flipped simultaneously. Thus, the move preserves energy while providing a mechanism for the system to traverse otherwise disconnected regions of the configuration space. Within PAMC, NCM is invoked periodically when population dynamics stall. By enabling larger collective moves, NCM diversifies the population and increases the likelihood of escaping plateaus. This contributes to better convergence properties and improved final cut values in Max-Cut optimization benchmarks.
    \item \textbf{Adaptive Temperature Control}: Another critical component of PAMC is the annealing schedule, which determines how the inverse temperature $\beta$ is updated during the search. Rather than relying entirely on a fixed linear or exponential schedule, PAMC uses an adaptive control rule based on the recent evolution of the best cut value in the population. Specifically, the algorithm monitors the number of distinct best-cut values observed over recent temperature windows to distinguish short-term stagnation from stronger stagnation. In the experiments reported here, we used the following rule. If the recent 25-step window contains at most three distinct best-cut values, while the recent 40-step window contains more than five, the algorithm identifies mild stagnation and reverses the inverse-temperature update, temporarily reheating the population to restore exploration. If both windows indicate stagnation, the algorithm applies a stronger intervention by reducing the inverse temperature by a factor of $0.8$ and activating the iso-cluster move. Once progress resumes, the algorithm returns to the usual cooling direction.
 This adaptive schedule therefore regulates the level of randomness according to the population’s optimization progress, balancing continued exploration with local refinement.
 
\end{itemize}

We examine how these enhancements influence the efficiency of landscape exploration within PAMC. In particular, we compare a conventional PAMC implementation with the augmented version that incorporates an adaptive temperature schedule and non-local cluster moves. This comparison allows us to isolate the effect of these enhancements on the population dynamics while holding the other algorithmic settings fixed.

To quantify the extent of landscape exploration, we monitored the diversity among replicas at intermediate stages of the algorithm using ${d}_{\mathrm H}$. This metric measures structural dissimilarity between candidate solutions. By tracking its evolution over time, we can determine whether the population rapidly collapses into a single basin or continues to explore multiple regions of the solution space. The results, shown in Fig. \ref{fig:g62_2x2}, reveal a clear difference between the two cases. As observed in Fig. \ref{fig:g62_2x2}(a), conventional PAMC quickly converges to a narrow region of the solution space, as indicated by the rapid decrease in ${d}_{\mathrm H}$. After this initial collapse, the replicas remain confined to the same basin until the end of the run, suggesting limited exploration capability.

In contrast, Fig. \ref{fig:g62_2x2}(b) shows that the augmented PAMC maintains a higher level of diversity throughout the run. The inclusion of nonlocal cluster moves enables occasional energy-preserving long-range transitions, increasing the likelihood that replicas escape locally confined regions. At the same time, the adaptive temperature controller introduces additional stochasticity by temporarily reheating the population. By reducing the inverse temperature $\beta$, this reheating step increases the likelihood of energetically unfavorable spin updates and further helps replicas move away from locally trapped configurations. As a result, even at later iterations of the annealing process, the ensemble retains a broad distribution of ${d}_{\mathrm H}$, typically in the range of (0.1) to (0.3). This suggests that multiple regions of the energy landscape continue to be explored during the later iterations of the run.

\subsection{Impact of PAMC Parameters}

\begin{figure}[h]
    \centering
    \includegraphics[width=1\linewidth]{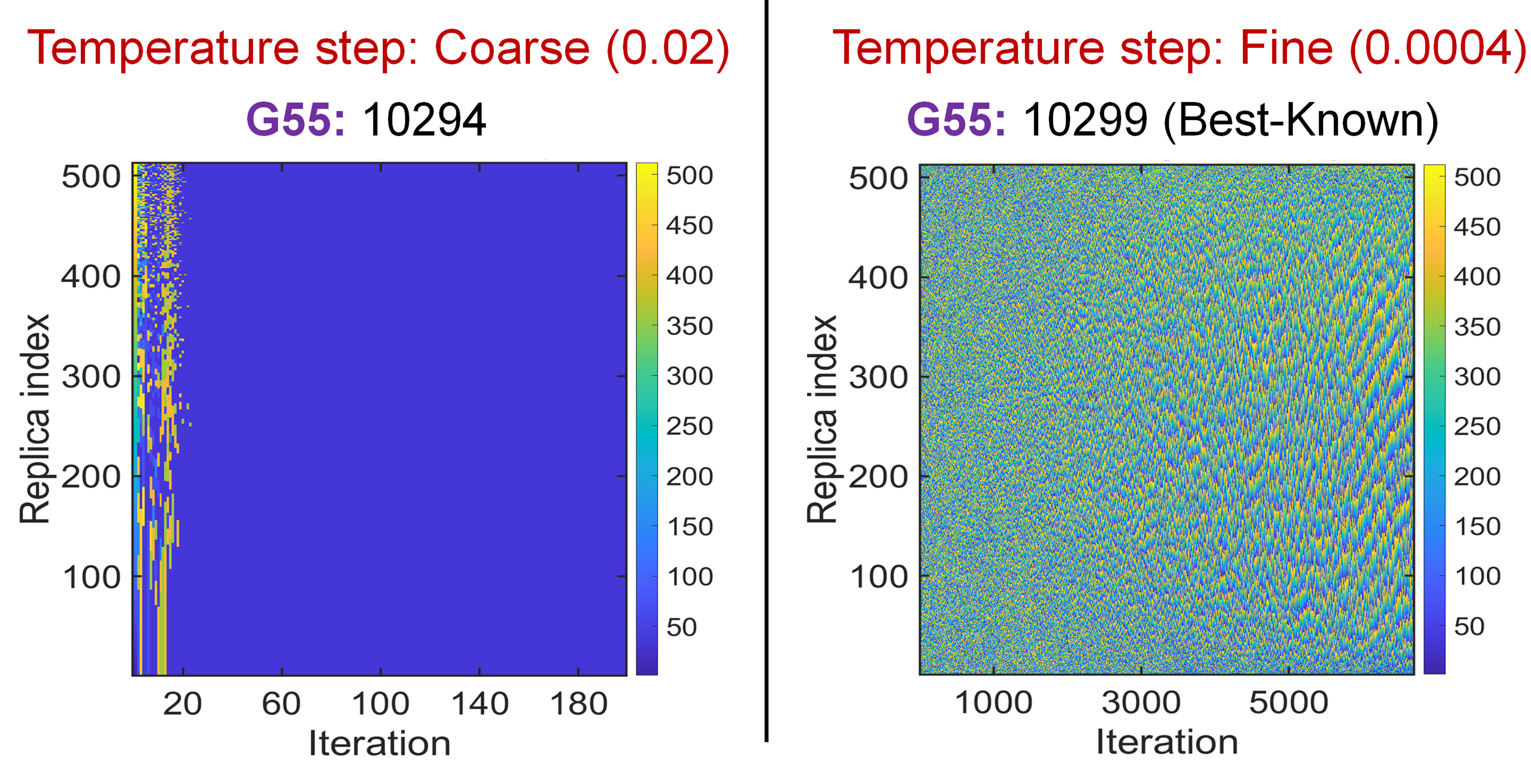}
    \caption{\justifying Genealogical evolution of the replica population ($R=512$) in PAMC for the G55 (N: 5000; E: 12498) Max-Cut instance under two different temperature-step sizes. Each color represents a replica, and the persistence or disappearance of colors indicates how resampling evolves the population during annealing.}
    \label{fig:lineage_plot}
\end{figure}

Although the pairwise Hamming distance captures the structural diversity among replicas, it does not fully explain how this diversity is shaped by the resampling mechanism. To better understand this process, we turn to lineage analysis, which provides a direct view of how replica families evolve and compete during the annealing process. The lineage plots in Fig. \ref{fig:lineage_plot} show that the effect of resampling in PAMC is strongly modulated by the temperature-step size and that the optimal schedule depends on the graph instance. In these plots, each color represents an ancestral lineage, so the persistence or disappearance of colors provides a direct view of how strongly resampling concentrates the population as the temperature is lowered. For G55 instance, the final Max-Cut solution changes noticeably with the temperature discretization, and the associated lineage plots indicate that coarse temperature steps lead to rapid genealogical collapse. In this regime, resampling becomes too strong, causing a few lineages to dominate early and eliminating alternative families before they can be sufficiently refined by the MCMC updates. As a result, the population loses diversity too quickly and is more likely to become trapped in suboptimal basins. By contrast, a finer temperature schedule mitigates this effect by producing a more gradual reduction in lineage diversity, thereby preserving exploration for longer, and allowing multiple families to compete over a wider range of temperatures, which results in an improved final Max-Cut solution. For this finer schedule, the number of MCMC sweeps per temperature stage was reduced from $(400)$ to $(10)$, since the smaller spacing between successive temperatures leads to a more gradual change in the target distribution and therefore requires less local re-equilibration at each step. The choice also avoids an excessive increase in total computational cost due to the larger number of temperature levels.\

These results show that temperature-step size acts as a practical control on the exploration-refinement balance in PAMC, such that coarse steps strengthen selection and can accelerate convergence on favorable landscapes, whereas fine steps preserve diversity and are more effective for rugged instances where extended competition among lineages is required. 
More broadly, these diagnostics provide insight into how PAMC interacts with the energy landscape induced by each graph instance and reveal how the structure of the instance affects the algorithmic dynamics.

\begin{figure}[h]
    \centering
    \includegraphics[width=0.7\linewidth]{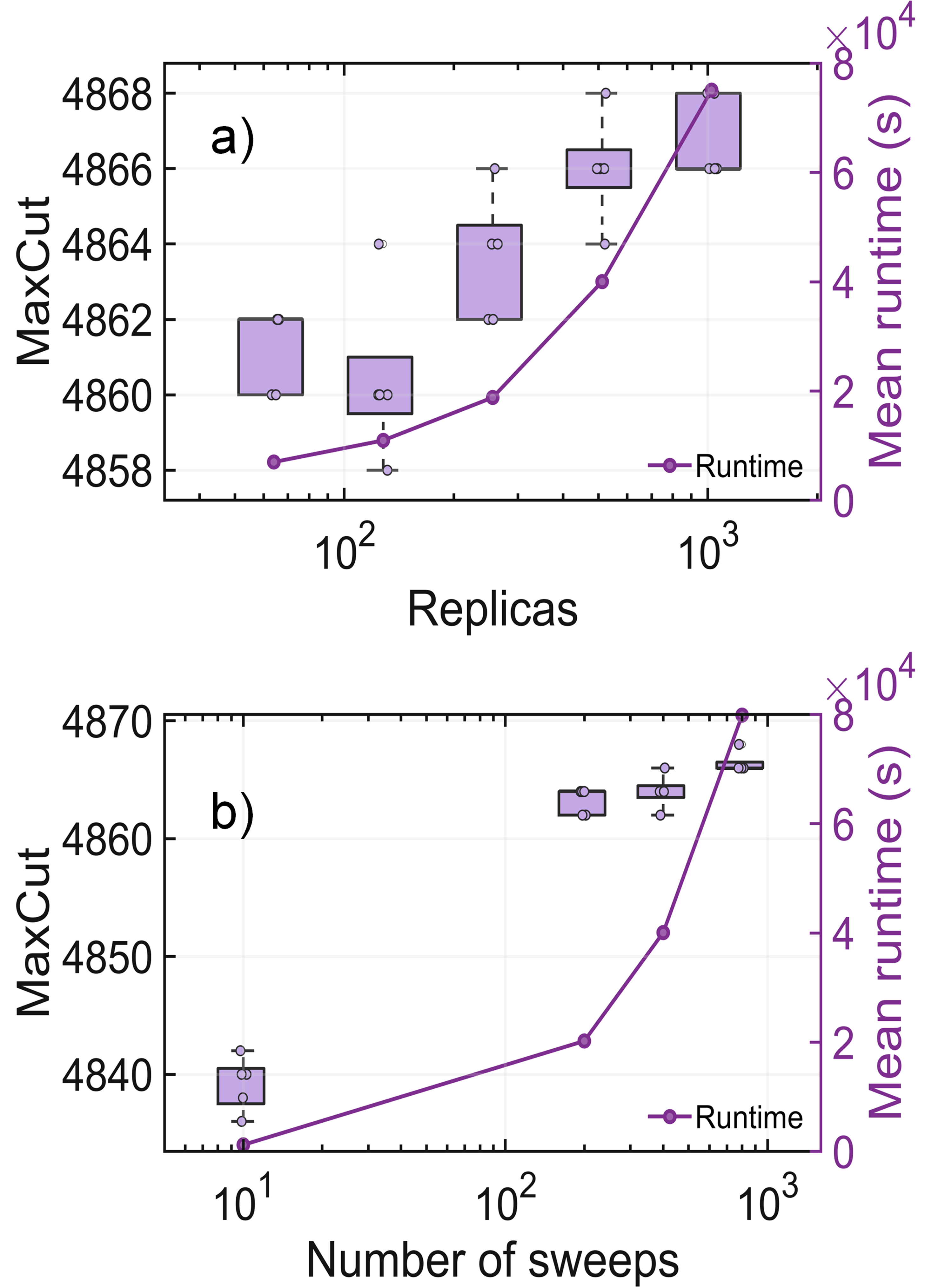}
    \caption{\justifying Effect of the number of replicas and MCMC sweeps on PAMC performance for the G62 (N: 7000; E: 14000) Max-Cut instance. (a) Max-Cut values obtained for different replica counts at sweep (=400). Mean runtime shown on the secondary axis. (b) Max-Cut values obtained for different numbers of sweeps at replica count (=512), with mean runtime shown on the secondary axis. }
    \label{replica_sweeps_Max-Cut}
\end{figure}

Fig. \ref{replica_sweeps_Max-Cut} examines the effect of two key computational parameters on PAMC performance for the G62 instance: the number of replicas and the number of MCMC sweeps, which directly control the breadth of population-level exploration and the depth of local refinement, respectively. In both cases, the results show that improved solution quality comes at an increased computational cost, but the gains are not proportional across the full parameter range.\
For Fig. \ref{replica_sweeps_Max-Cut}(a), increasing the number of replicas generally improves the Max-Cut value, indicating that a larger population enhances the ability of PAMC to sample and preserve promising regions of the configuration space. With more replicas, the algorithm can maintain broader coverage of the landscape during the annealing process, which increases the probability that high-quality basins are discovered and retained through resampling. This trend is visible from (64) to (512) replicas, where the distribution of Max-Cut values shifts upward. However, the improvement becomes much smaller between (512) and (1024) replicas, while the mean runtime increases sharply. These results indicate a regime of diminishing returns: beyond a certain population size, additional replicas substantially increase computational cost while yielding only marginal gains in solution quality. Thus, increasing the replica count improves exploration up to a point, but further increases may not be cost-effective unless the goal is to maximize solution quality regardless of runtime.\\
\indent For Fig. \ref{replica_sweeps_Max-Cut}(b), increasing the number of sweeps also improves the Max-Cut value initially, showing that local refinement through additional MCMC updates helps the replicas better exploit the basins they occupy. In particular, moving from very few sweeps to a moderate number produces a clear improvement, indicating that insufficient local equilibration can limit performance. However, beyond this range, the Max-Cut values begin to plateau, while the runtime continues to rise substantially. The results therefore show that simply increasing the amount of local refinement does not guarantee continued improvement in solution quality. Once the replicas have already been refined adequately within their current basins, additional sweeps mostly increase computational cost without producing comparable gains. This again highlights a diminishing-return effect, now with respect to local search effort rather than population size.\\
\indent Overall, the figure shows that both population size and local refinement depth must be balanced carefully, and neither parameter should be increased indefinitely. For the G62 instance, the results indicate that moderate-to-large values are sufficient to obtain most of the achievable gain, while very large values primarily increase runtime. Therefore, the result highlights the practical trade-off between solution quality and computational cost, and suggests that efficient PAMC performance depends on identifying a parameter regime in which the marginal improvement in Max-Cut remains in proportion with the additional time and resource investment.

\subsection{Max-Cut}

\begin{table*}[t]
\centering
\caption{Comparison of the time-to-solution (TTS) of SBM and PAMC on selected G-set benchmark instances, using the solution reported by SBM as target solution quality. For each graph, the table lists the best-known cut value, the SBM solution used as the target, and the corresponding TTS and success probability $P_s$ for both SBM and PAMC. Statistics are computed over 100 independent runs. Graphs shown in parentheses were evaluated using only 10 independent runs. The PAMC speedup is computed as the ratio of the SBM TTS to the PAMC TTS. Instances in which PAMC achieves a lower TTS than SBM for the same target solution quality are highlighted in bold. Entries marked as ``--'' correspond to instances for which PAMC was unable to reach the target solution quality with the specified parameter settings across all runs.}
\label{tab:sbm-pamc}
\scriptsize
\setlength{\tabcolsep}{3.8pt}
\renewcommand{\arraystretch}{1.1}
\begin{tabular}{c c c c||c@{\hspace{6pt}}c|c@{\hspace{6pt}}c||c}
\toprule
\multirow{2}{*}{Graph} & \multirow{2}{*}{$n$} & \multirow{2}{*}{Best-known} & \multirow{2}{*}{SBM Sol.} & \multicolumn{2}{c|}{SBM \cite{Goto2021highperformance}} & \multicolumn{2}{c||}{PAMC} & \multirow{2}{*}{Speedup} \\
\cmidrule(lr){5-6} \cmidrule(lr){7-8}
 &  &  &  & TTS (s) & $P_s$ & TTS (s) & $P_s$ &  \\
\midrule
G35 & 2000  & 7687  & 7686  & 8319  & 0.0005 & \textbf{348.30}    & 0.60 & \textbf{23.88$\times$} \\
G36 & 2000  & 7680  & 7680  & 62647 & 0.0001 & \textbf{12361.42}  & 0.02 & \textbf{5.07$\times$} \\
G37 & 2000  & 7691  & 7691  & 27343 & 0.0002 & \textbf{11244.34}  & 0.02 & \textbf{2.43$\times$} \\
G58 & 5000  & 19294 & 19283 & 21638 & 0.0010 & \textbf{9062.50}   & 0.93 & \textbf{2.39$\times$} \\
G59 & 5000  & 6088  & 6086  & 22651 & 0.0010 & -                  & -    & - \\
G60 & 7000  & 14190 & 14186 & 21827 & 0.0010 & 25798.32           & 0.51 & 0.85$\times$ \\
G61 & 7000  & 5798  & 5796  & 10883 & 0.0020 & -                  & -    & - \\
G62 & 7000  & 4872  & 4862  & 20501 & 0.0010 & 20931.07           & 0.81 & 0.98$\times$ \\
G63 & 7000  & 27045 & 27023 & 31279 & 0.0010 & \textbf{5345.05}   & 0.96 & \textbf{5.85$\times$} \\
G64 & 7000  & 8752  & 8739  & 31448 & 0.0010 & 75175.96           & 0.15 & 0.42$\times$ \\
G66 & 9000  & 6364  & 6342  & 27408 & 0.0010 & \textbf{7099.79}   & 0.99 & \textbf{3.86$\times$} \\
G70 & 10000 & 9595  & 9578  & 31599 & 0.0010 & \textbf{12804.60}  & 0.99 & \textbf{2.47$\times$} \\
(G77) & 14000 & 9940  & 9904  & 46760 & 0.0010 & \textbf{22225}     & 1.00 & \textbf{2.10$\times$} \\
(G81) & 20000 & 14060 & 13392 & 62194 & 0.0010 & \textbf{37397}     & 1.00 & \textbf{1.66$\times$} \\
\bottomrule
\end{tabular}
\end{table*}

The augmented PAMC solver was benchmarked against SBM by evaluating two key metrics: a) Time-to-solution (TTS) and b) Solution quality. The SBM provides a stringent reference point because it has reported state-of-the-art time-to-solution performance on the same G-set benchmark suite, making it a strong baseline for evaluating the runtime efficiency of the proposed PAMC framework~\cite{Goto2021highperformance}.
TTS is the total time required for a solver to reach a target solution with 99\% success probability and is expressed as;
\begin{equation}
    \text{TTS} = T_{\text{com}} \cdot 
    \frac{\log(1-0.99)}{\log(1-P_s)},
\end{equation}
where $T_{\text{com}}$ is the total compute time per run,
and $P_s$ is the empirical success probability.

\begin{table*}[t]
\centering
\caption{Comparison of solution quality between SBM and PAMC on selected G-set benchmark instances, evaluated at a target runtime equal to SBM’s time-to-solution (TTS). The selected instances are those for which SBM reports a TTS greater than 7000 s. For each graph, the table reports the best-known cut value, SBM’s TTS, the solution obtained by SBM, the solution obtained by PAMC with the same target time, and the best overall solution.}
\label{tab:sbm-pamc-best}
\scriptsize
\setlength{\tabcolsep}{4.5pt}
\renewcommand{\arraystretch}{1.1}
\begin{tabular}{c c c c||c|c||c}
\toprule
Graph & $n$ & Best-known & SBM TTS & SBM Sol. & PAMC Sol. & Best Sol. Achieved  \\
\midrule
G35 & 2000  & 7687  & 8319  & 7686  & 7686  & \textbf{7687} \\
G36 & 2000  & 7680  & 62647  & 7680  & 7680  & 7680 \\
G37 & 2000  & 7691  & 27343  & 7691  & 7691  & 7691 \\
G58 & 5000  & 19294 & 21638 & 19283 & \textbf{19286} & \textbf{19293} \\
G59 & 5000  & 6088  & 22651 & 6086  &  6073    &   6080           \\
G60 & 7000  & 14190 & 21827 & 14186 & \textbf{14186} & \textbf{ 14188} \\
G61 & 7000  & 5798  & 10883 & 5796  & 5791  & \textbf{5798} \\
G62 & 7000  & 4872  & 20501 & 4862  & 4862  & \textbf{4866} \\
G63 & 7000  & 27045 & 31279 & 27023 & \textbf{27031} & \textcolor{blue}{\textbf{27047\textcolor{blue}{*}}} \\
G64 & 7000  & 8752  & 31448 & 8739  & 8732  & \textbf{8750} \\
G66 & 9000  & 6364  & 27408 & 6342  & \textbf{6346}  & \textbf{6356} \\
G70 & 10000 & 9595  & 31599 & 9578  & \textbf{9586}  & \textbf{9592} \\
G77 & 14000 & 9940  & 46760 & 9904  & \textbf{9908}  & \textbf{9918} \\
G81 & 20000 & 14060  & 62194 & 13392  & \textbf{14008}  & \textbf{14014} \\
\bottomrule
\end{tabular}
\end{table*}

The first experiment evaluated the TTS required by PAMC to reach the Max-Cut value reported by SBM for each benchmark instance. 

For each graph, the target solution was fixed to SBM's reported best cut value. We launched $100$ independent PAMC runs and terminated each run as soon as the target solution was reached. The wall-clock time required to reach the target was recorded as the total compute runtime $T_{\text{com}}$ for that run. 

For each candidate runtime ${T_{\text{com}}}$, the empirical success probability $P_s(T_{com})$ was calculated as the fraction of the $100$ independent runs that reached the target (SBM best cut) within that time. TTS was then computed for each candidate pair $\lbrace T_{\text{com}}, P_s(T_{\text{com}})\rbrace$, and the lowest TTS value obtained over this scan was reported.

Table~\ref{tab:sbm-pamc} presents the TTS computed for PAMC with those reported for SBM. In the main text, we focus on the harder benchmark instances for which SBM reports TTS greater than 7000 s. This selection emphasizes graphs where the runtime is governed mainly by the difficulty of exploring the optimization landscape. This is important because our current GPU implementation is not yet performance-optimized, and hence smaller and easier graphs are reflective of implementation costs more than algorithmic efficiency. Nevertheless, detailed results for all the instances have been presented in Supplementary Note \ref{detailed_results}. The results indicate that on several larger and more challenging graphs, PAMC attains the same cut value as SBM in a comparatively shorter runtime as shown in Table \ref{tab:sbm-pamc}.

The second benchmark experiment evaluated solution quality under a fixed runtime budget, using SBM’s reported TTS as the reference wall-clock time for each instance.
To do so, we compared the solutions obtained by PAMC with SBM's, while allowing the PAMC solver a runtime budget $T_{com}$ approximately equal to the TTS reported by SBM. 
Each graph was executed 10 times. For each graph, the best solution achieved across 10 runs was recorded. The best solutions achieved across these runs are shown in Table~\ref{tab:sbm-pamc-best} (Best Sol. Achieved). The comparable solutions are those generated with 100\% probability across all runs also shown in Table \ref{tab:sbm-pamc-best} (PAMC Sol.). A total of 6 out of 14 graphs achieved better solutions than SBM, with 5 graphs generating the same solution as SBM when run for an equal amount of time.  On the other hand, when considering only the best solutions achieved across the 10 runs, PAMC equaled or outperformed SBM in all the graphs considered in Table~\ref{tab:sbm-pamc-best} except one. The solver also found a new best-known solution for the G63 instance. A spin configuration corresponding to the best-known Max-Cut for the G63 instance can be found at~\cite{khan2025new}.

\subsection{Max-$3$-Cut}

As noted earlier, approaches based on binary Ising formulations naturally support Max-Cut but require an additional mapping step for Max-$K$-Cut with ($K>2$). This mapping introduces auxiliary variables and inflates the effective problem size, which can affect time-to-solution, solution quality, and energy efficiency in both physical hardware and digital emulators. To avoid this overhead, we adopt the (q)-state Potts model, a standard statistical-physics generalization of the Ising model in which each spin can occupy one of (q) discrete states \cite{potts1952generalized, wu1982potts}. For Max-$K$-Cut, we set (q=$K$), so each graph vertex is represented directly by a single $K$-state Potts spin. Thus, no graph-conversion step or auxiliary binary expansion is required, and the problem size remains equal to the original input graph. Here, we focus on the Max-$3$-Cut case as a representative multi-state benchmark, although the same Potts-based formulation is applicable to Max-$K$-Cut with $K>3$.

 \begin{table}[h]
\centering
\caption{\justifying Comparative Max-$3$-Cut (M$3$C) results for G-set instances on which our solver improves the previously best-known Max-$3$-Cut solution. $\Delta$ = $M3C_{\text{Our-Work}} - M3C_{\text{Best-Known}}$; `--' indicates that the solution to the instance was not reported in the work; MOH: Multi-Operator Heuristic; $n$: nodes; $m$: edges. Entries marked with ($^*$) were obtained using the second update strategy, based on cascaded p-bits for Gibbs sampling.}
\label{tab:max3cut-results}
\scriptsize
\setlength{\tabcolsep}{3pt}
\begin{tabular}{r r r r rr r r}
\toprule
\multirow{2}{*}{Graph} & \multirow{2}{*}{$n$} & \multirow{2}{*}{$m$} & \multirow{2}{*}{MOH \cite{Ma2017MOH}} & \multicolumn{2}{c}{PLS \cite{garvardt2024parameterized}} & \multirow{2}{*}{Our} & \multirow{2}{*}{$\Delta$} \\
\cmidrule(lr){5-6}
 &  &  &  & LS & ILP &  &  \\
\midrule
G16 & 800 & 4672 & 3991 & -- & -- & \textbf{3992} & 1 \\
G17 & 800 & 4667 & 3983 & -- & -- & \textbf{3986} & 3 \\
G21 & 800 & 4667 & 1109 & -- & -- & \textbf{1110} & 1 \\
G22 & 2000 & 19990 & 17167 & -- & -- & \textbf{17169} & 2 \\
G23 & 2000 & 19990 & 17168 & -- & -- & \textbf{17172} & 4 \\
G25 & 2000 & 19990 & 17163 & 17164 & -- & \textbf{17167} & 3 \\
G26 & 2000 & 19990 & 17154 & 17155 & -- & \textbf{17157} & 2 \\
G28 & 2000 & 19990 & 3973 & 3975 & -- & \textbf{3978} & 3 \\
G29 & 2000 & 19990 & 4106 & -- & -- & \textbf{4110} & 4 \\
G30 & 2000 & 19990 & 4119 & -- & -- & \textbf{4121} & 2 \\
G31 & 2000 & 19990 & 4003 & 4005 & -- & \textbf{4008} & 3 \\
G35 & 2000 & 11778 & 10046 & 10048 & -- & \textbf{10058} & 10 \\
G36 & 2000 & 11766 & 10039 & -- & -- & \textbf{10047} & 8 \\
G37 & 2000 & 11785 & 10052 & 10053 & 10053 & \textbf{10058} & 5 \\
G38 & 2000 & 11779 & 10040 & -- & -- & \textbf{10054} & 14 \\
G39 & 2000 & 11778 & 2903 & -- & -- & \textbf{2910} & 7 \\
G40 & 2000 & 11766 & 2870 & 2871 & -- & \textbf{2878} & 7 \\
G41 & 2000 & 11785 & 2887 & 2888 & -- & \textbf{2889} & 1 \\
G42 & 2000 & 11779 & 2980 & -- & -- & \textbf{2983} & 3 \\
G48 & 3000 & 6000 & 6000 & -- & -- & 6000* & 0 \\
G49 & 3000 & 6000 & 6000 & -- & -- & 6000* & 0 \\
G50 & 3000 & 6000 & 6000 & -- & -- & 6000* & 0 \\
G51 & 1000 & 5909 & 5037 & -- & -- & \textbf{5039} & 2 \\
G52 & 1000 & 5916 & 5040 & -- & -- & \textbf{5043} & 3 \\
G53 & 1000 & 5914 & 5039 & -- & -- & \textbf{5042} & 3 \\
G54 & 1000 & 5916 & 5036 & -- & -- & \textbf{5040} & 4 \\
G56 & 5000 & 12498 & 4752 & 4757 & -- & \textbf{4763} & 6 \\
G57 & 5000 & 10000 & 4083 & 4092 & 4103 & \textbf{4114} & 11 \\
G58 & 5000 & 29570 & 25195 & -- & -- & \textbf{25216} & 21 \\
G59 & 5000 & 29570 & 7262 & 7276 & -- & \textbf{7309} & 33 \\
G60 & 7000 & 17148 & 17076 & -- & -- & \textbf{17082} & 6 \\
G61 & 7000 & 17148 & 6853 & 6861 & -- & \textbf{6872} & 11 \\
G62 & 7000 & 14000 & 5685 & 5710 & 5706 & \textbf{5731} & 21 \\
G63 & 7000 & 41459 & 35322 & 35318 & -- & \textbf{35355} & 33 \\
G64 & 7000 & 41459 & 10443 & 10437 & -- & \textbf{10508} & 65 \\
G65 & 8000 & 16000 & 6490 & 6512 & 6535 & \textbf{6544} & 9 \\
G66 & 9000 & 18000 & 7416 & 7442 & 7443 & \textbf{7475} & 32 \\
G67 & 10000 & 20000 & 8086 & 8116 & 8141 & \textbf{8152} & 11 \\
G70 & 10000 & 9999 & 9999 & -- & -- & 9999* & 0 \\
G72 & 10000 & 20000 & 8192 & 8244 & 8244 & \textbf{8259} & 15 \\
\bottomrule
\end{tabular}
\end{table}

For the Max-\(K\)-Cut problem, each vertex is assigned a Potts spin
\(\sigma_i \in \{1,2,\dots,K\}\), allowing direct optimization over \(K\) partitions
without reducing the problem to a binary Ising form. The energy of a
configuration is defined as
\begin{equation}
    E(\boldsymbol{\sigma}) = - \sum_{i,j} J_{ij}\bigl(2\delta(\sigma_i,\sigma_j)-1\bigr),
\end{equation}
where \(\delta(.)\) is the Kronecker delta. In this formulation, the coupling \(J_{ij}\) is edge dependent but not label dependent, which is appropriate for Max-$K$-Cut because all partitions are equivalent and the objective depends only on whether adjacent vertices share the same label. This form penalizes assignments in which adjacent vertices occupy the same partition and favors assignments in different partitions.
For each vertex \(i\), a $k$-label local field is formed,
\begin{equation}
    \mathrm{syn}_k(i) = \sum_j J_{ij}\,\delta(\sigma_j,k),
    \qquad k=1,\dots,K,
\end{equation}
which measures the weighted influence of neighbors currently occupying state
$k$. 

We then update the state of the vertex \(i\) using two complementary methods. The first method is a randomized softmax-based K-state spin update, in which the $K$-label local field is converted into a softmax distribution over the possible spin states.
For \(k=1,\dots,K\), the conditional softmax probability is
\begin{equation}
    P(\sigma_i = k \mid \{\sigma_j\}) =
    \frac{\exp\!\left(\beta\,\mathrm{syn}_k(i)\right)}
    {\sum_{m=1}^{K} \exp\!\left(\beta\,\mathrm{syn}_m(i)\right)} .
\end{equation}
The resulting probabilities are then used in a stochastic update rule. Specifically, each probability is perturbed by independent uniform noise, and the new state is selected as
\begin{equation}
\sigma_i \leftarrow \arg\max_{k}
\left[
P(\sigma_i=k \mid {\sigma_j}) - \gamma U_k
\right],
\end{equation}
where \(U_k \sim {U}(0,1)\) and \(\gamma\) controls the noise amplitude. 

The second method is an exact cascaded binary p-bit sampler derived specifically for the ($K=3$) case. Rather than sampling directly from a three-state softmax distribution, this method decomposes the same three-state Gibbs update into two sequential binary stochastic decisions. The cascade reproduces the exact three-state Potts conditional distribution while using only binary probabilistic updates, providing a binary-compatible formulation of the Max-3-Cut Gibbs sampler. The full derivation of the cascaded sampler is provided in supplementary material~\ref{derivation_potts}. Although the cascaded construction exactly reproduces the three-state Gibbs conditional distribution, our empirical results in Table~\ref{tab:max3cut-results} indicate that, in the present GPU implementation, the direct softmax update is more efficient for most benchmark instances. An exception is observed for the tripartite G-set instances, namely G48, G49, G50, and G70, where the cascaded binary sampler performs competitively or favorably. Understanding this instance-dependent behavior, and whether binary-compatible Potts sampling can be further optimized for structured graph classes, will be investigated in future work.

For each graph instance, the algorithm was run $10$ times independently, and the best solution obtained across these runs is reported for these $36$ graph instances on which the solver establishes new best-known Max-3-Cut solutions. Extended Max-3-Cut results are provided in Supplementary Material \ref{detailed_results}.

\subsection{Scalability on a Large-Scale Ising Problem}

\begin{figure}[h]
    \centering
    \includegraphics[width=1\linewidth]{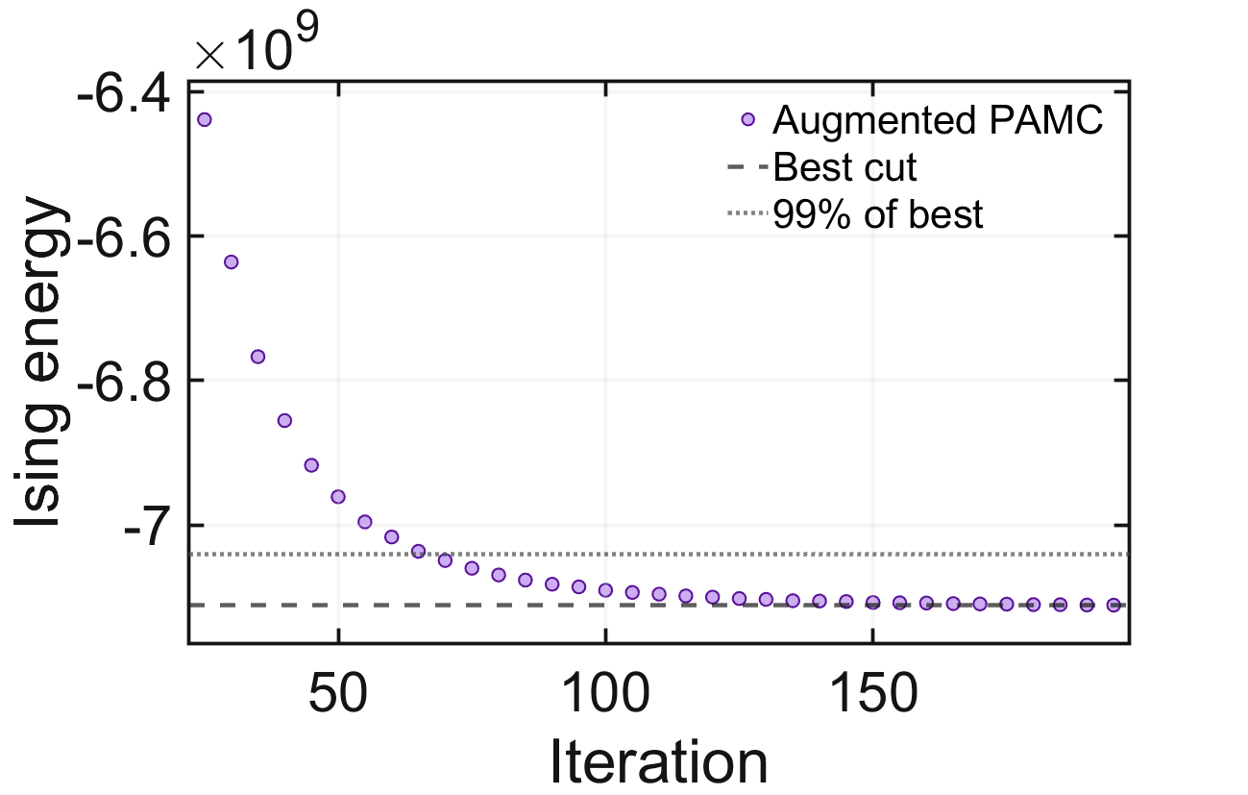}
    \caption{\justifying Scalability of the augmented PAMC solver on an ultra-large fully connected Ising problem with $N=100,000$ spins. }
    \label{ultrasscale}
\end{figure}

Inspired by SBM results, we also evaluate the scalability of the proposed augmented PAMC solver beyond the Gset graph sizes. We considered a large, fully connected Ising instance with $N=100,000$ spins and signed weights. The instance was generated using the pseudocode described in ~\cite{Goto2021highperformance}. Since the graph is fully connected, it represents a substantially larger and complementary dense Ising instance than the sparse G-set graphs considered above. 

For this problem, the expected ground state energy was estimated using the Parisi-limit expression also reported in \cite{Goto2021highperformance}. The augmented PAMC solver reached the estimated target cut value in approximately $10^4$ seconds, as shown in Fig. \ref{ultrasscale}. This result suggests that the proposed framework can be extended to large dense Ising instances.

\section{Conclusion}
This work presented an augmented Population Annealing Monte Carlo framework for large-scale combinatorial optimization. The method combines population resampling with stagnation-driven temperature control and energy-preserving nonlocal cluster moves, allowing the replica population to concentrate around promising solution regions while retaining sufficient diversity to avoid premature convergence.

The framework was implemented on a GPU and applied to both Max-Cut and Max-$K$-Cut within a common computational architecture. On G-set Max-Cut benchmarks, the augmented PAMC solver achieves competitive time-to-solution relative to the simulated bifurcation machine, improves the best-known solution for the G63 instance, and matches or improves solution quality on most tested large instances under comparable runtime budgets. For Max-$3$-Cut, the solver establishes new best-known solutions on 36 G-set instances, demonstrating that the approach extends beyond binary Ising formulations.

The solver also scales to a fully connected 100{,}000-spin Ising instance. Taken together, these results show that feedback-controlled population dynamics provide an effective and GPU-compatible strategy for steering stochastic search in large, complex energy landscapes.

\section*{Acknowledgments}
The authors gratefully acknowledge Prof. Kerem Camsari for his valuable input. This work was supported by the National Science Foundation grant \# 2132198. The authors also acknowledge support from a DAC Analytics Resource Award from the University of Virginia.



\clearpage
\bibliography{references}

\begin{thebibliography}{45}

\bibitem{seo2015edgeset}
K.~Seo, S.~Hyun, and Y.-H. Kim.
\newblock An edge-set representation based on a spanning tree for searching cut space.
\newblock \emph{IEEE Transactions on Evolutionary Computation}, 19(4):465--473, 2015.

\bibitem{benlic2011multilevel}
U.~Benlic and J.-K. Hao.
\newblock A multilevel memetic approach for improving graph k-partitions.
\newblock \emph{IEEE Transactions on Evolutionary Computation}, 15(5):624--642, 2011.

\bibitem{Metropolis1953equation}
N.~Metropolis, A.~W. Rosenbluth, M.~N. Rosenbluth, A.~H. Teller, and E.~Teller.
\newblock Equation of state calculations by fast computing machines.
\newblock \emph{The Journal of Chemical Physics}, 21(6):1087--1092, 1953.

\bibitem{Kirkpatrick1983optimization}
S.~Kirkpatrick, C.~D. Gelatt, and M.~P. Vecchi.
\newblock Optimization by simulated annealing.
\newblock \emph{Science}, 220(4598):671--680, 1983.

\bibitem{Iba2001population}
Y.~Iba.
\newblock Population Monte Carlo algorithms.
\newblock \emph{Transactions of the Japanese Society for Artificial Intelligence}, 16(2):279--286, 2001.

\bibitem{Earl2005parallel}
D.~J. Earl and M.~W. Deem.
\newblock Parallel tempering: Theory, applications, and new perspectives.
\newblock \emph{Physical Chemistry Chemical Physics}, 7(23):3910--3916, 2005.

\bibitem{Hukushima2003population}
K.~Hukushima and Y.~Iba.
\newblock Population annealing and its application to a spin glass.
\newblock In \emph{The Monte Carlo Method in the Physical Sciences: Celebrating the 50th Anniversary of the Metropolis Algorithm}, volume 690 of \emph{AIP Conference Proceedings}, pages 200--206. American Institute of Physics, 2003.

\bibitem{Huang2024ising}
K.-P. Huang, C.-F. Nien, Y.-T. Zhang, C.-K. Lee, and Y.-C. Wang.
\newblock GPU-based Ising machine for solving combinatorial optimization problems with enhanced parallel tempering techniques.
\newblock In \emph{2024 IEEE Asia Pacific Conference on Circuits and Systems (APCCAS)}, pages 636--640, Taipei, Taiwan, 2024. IEEE.

\bibitem{zhu2020borealis}
Z.~Zhu, C.~Fang, and H.~G. Katzgraber.
\newblock borealis---A generalized global update algorithm for Boolean optimization problems.
\newblock \emph{Optimization Letters}, 14(8):2495--2514, 2020.

\bibitem{delacourKerem2025two}
C.~Delacour, M.~M.~H. Sajeeb, J.~P. Hespanha, and K.~Y. Camsari.
\newblock Two-dimensional parallel tempering for constrained optimization.
\newblock \emph{Physical Review E}, 112(2):L023301, 2025.

\bibitem{nikhar2024allKerem}
S.~Nikhar, S.~Kannan, N.~A. Aadit, S.~Chowdhury, and K.~Y. Camsari.
\newblock All-to-all reconfigurability with sparse and higher-order Ising machines.
\newblock \emph{Nature Communications}, 15(1):8977, 2024.

\bibitem{Wang2015comparing}
W.~Wang, J.~Machta, and H.~G. Katzgraber.
\newblock Comparing Monte Carlo methods for finding ground states of Ising spin glasses: Population annealing, simulated annealing, and parallel tempering.
\newblock \emph{Physical Review E}, 92(1):013303, 2015.

\bibitem{Amey2018analysis}
C.~Amey and J.~Machta.
\newblock Analysis and optimization of population annealing.
\newblock \emph{Physical Review E}, 97(3):033301, 2018.

\bibitem{Barash2017gpu}
L.~Yu. Barash, M.~Weigel, M.~Borovsk\'y, W.~Janke, and L.~N. Shchur.
\newblock GPU accelerated population annealing algorithm.
\newblock \emph{Computer Physics Communications}, 220:341--350, 2017.

\bibitem{barzegar2024optimal}
A.~Barzegar, F.~Hamze, C.~Amey, and J.~Machta.
\newblock Optimal schedules for annealing algorithms.
\newblock \emph{Physical Review E}, 109(6):065301, 2024.

\bibitem{barzegar2018optimization}
A.~Barzegar, C.~Pattison, W.~Wang, and H.~G. Katzgraber.
\newblock Optimization of population annealing Monte Carlo for large-scale spin-glass simulations.
\newblock \emph{Physical Review E}, 98(5):053308, 2018.

\bibitem{ebert2022weighted}
P.~L. Ebert, D.~Gessert, and M.~Weigel.
\newblock Weighted averages in population annealing: Analysis and general framework.
\newblock \emph{Physical Review E}, 106(4):045303, 2022.

\bibitem{weigel2021understanding}
M.~Weigel, L.~Barash, L.~Shchur, and W.~Janke.
\newblock Understanding population annealing Monte Carlo simulations.
\newblock \emph{Physical Review E}, 103(5):053301, 2021.

\bibitem{Cirauqui2024population}
D.~Cirauqui, M.~\'A. Garc\'ia-March, J.~R. Mart\'inez Saavedra, M.~Lewenstein, and P.~R. Grzybowski.
\newblock Population annealing with topological defect driven nonlocal updates for spin systems with quenched disorder.
\newblock \emph{Physical Review B}, 109(14):144202, 2024.

\bibitem{Ingber1989VFSR}
L.~Ingber.
\newblock Very fast simulated re-annealing.
\newblock \emph{Mathematical and Computer Modelling}, 12(8):967--973, 1989.

\bibitem{Karabin2020adaptive}
M.~Karabin and S.~J. Stuart.
\newblock Simulated annealing with adaptive cooling rates.
\newblock \emph{The Journal of Chemical Physics}, 153(11):114103, 2020.

\bibitem{Katzgraber2006feedback}
H.~G. Katzgraber, S.~Trebst, D.~A. Huse, and M.~Troyer.
\newblock Feedback-optimized parallel tempering Monte Carlo.
\newblock \emph{Journal of Statistical Mechanics: Theory and Experiment}, 2006(03):P03018, 2006.

\bibitem{Miasojedow2013adaptive}
B.~Miasojedow, E.~Moulines, and M.~Vihola.
\newblock An adaptive parallel tempering algorithm.
\newblock \emph{Journal of Computational and Graphical Statistics}, 22(3):649--664, 2013.

\bibitem{boese1994best}
K.~D. Boese and A.~B. Kahng.
\newblock Best-so-far vs. where-you-are: implications for optimal finite-time annealing.
\newblock \emph{Systems \& Control Letters}, 22(1):71--78, 1994.

\bibitem{Houdayer2001cluster}
J.~Houdayer.
\newblock A cluster Monte Carlo algorithm for 2-dimensional spin glasses.
\newblock \emph{The European Physical Journal B}, 22(4):479--484, 2001.

\bibitem{Zhu2015isoenergetic}
Z.~Zhu, A.~J. Ochoa, and H.~G. Katzgraber.
\newblock Efficient cluster algorithm for spin glasses in any space dimension.
\newblock \emph{Physical Review Letters}, 115(7):077201, 2015.

\bibitem{Benlic2013BLS}
U.~Benlic and J.-K. Hao.
\newblock Breakout local search for the Max-Cut problem.
\newblock \emph{Engineering Applications of Artificial Intelligence}, 26(3):1162--1173, 2013.

\bibitem{Shylo2015teams}
V.~P. Shylo, F.~Glover, and I.~V. Sergienko.
\newblock Teams of global equilibrium search algorithms for solving the weighted maximum cut problem in parallel.
\newblock \emph{Cybernetics and Systems Analysis}, 51(1):16--24, 2015.

\bibitem{Goudet2024island}
O.~Goudet, A.~Go\"effon, and J.-K. Hao.
\newblock A large population island framework for the unconstrained binary quadratic problem.
\newblock \emph{Computers \& Operations Research}, 168:106684, 2024.

\bibitem{Ma2017MOH}
F.~Ma and J.-K. Hao.
\newblock A multiple search operator heuristic for the max-$k$-cut problem.
\newblock \emph{Annals of Operations Research}, 248(1):365--403, 2017.

\bibitem{garvardt2024parameterized}
J.~Garvardt, N.~Gr\"uttemeier, C.~Komusiewicz, and N.~Morawietz.
\newblock Parameterized local search for Max $c$-Cut.
\newblock \emph{arXiv preprint} arXiv:2409.13380, 2024.

\bibitem{Goto2021highperformance}
H.~Goto, K.~Endo, M.~Suzuki, Y.~Sakai, T.~Kanao, Y.~Hamakawa, R.~Hidaka, M.~Yamasaki, and K.~Tatsumura.
\newblock High-performance combinatorial optimization based on classical mechanics.
\newblock \emph{Science Advances}, 7(6):eabe7953, 2021.

\bibitem{zick2026cosm}
K.~M. Zick, N.~Shukla, and A.~Marakov.
\newblock Cosm: Collective switched motion for fast and accurate sparse Ising optimization.
\newblock \emph{arXiv preprint} arXiv:2605.30355, 2026.

\bibitem{Camsari2017pbits}
K.~Y. Camsari, R.~Faria, B.~M. Sutton, and S.~Datta.
\newblock Stochastic $p$-bits for invertible logic.
\newblock \emph{Physical Review X}, 7(3):031014, 2017.

\bibitem{Aadit2022massively}
N.~A. Aadit, A.~Grimaldi, M.~Carpentieri, L.~Theogarajan, J.~M. Martinis, G.~Finocchio, and K.~Y. Camsari.
\newblock Massively parallel probabilistic computing with sparse Ising machines.
\newblock \emph{Nature Electronics}, 5(7):460--468, 2022.

\bibitem{Borders2019factorization}
W.~A. Borders, A.~Z. Pervaiz, S.~Fukami, K.~Y. Camsari, H.~Ohno, and S.~Datta.
\newblock Integer factorization using stochastic magnetic tunnel junctions.
\newblock \emph{Nature}, 573(7774):390--393, 2019.

\bibitem{yang2025250}
S.~Yang, A.~Grimaldi, Y.~Bao, E.~Raimondo, J.~Si, G.~Finocchio, and H.~Yang.
\newblock 250 magnetic tunnel junctions-based probabilistic Ising machine.
\newblock \emph{arXiv preprint} arXiv:2506.14590, 2025.

\bibitem{Bashar2024kstate}
M.~K. Bashar, A.~Hasan, and N.~Shukla.
\newblock Designing a K-state p-bit engine.
\newblock \emph{arXiv preprint} arXiv:2403.06436, 2024.

\bibitem{Cheong2026potts}
S.~Cheong, S.~H. Lee, J.~Han, J.-Y. Park, D.~H. Shin, Y.~H. Jang, S.~K. Shim, S.~Kim, C.~S. Hwang, and J.-K. Han.
\newblock Multi-state probabilistic computing using floating-body MOSFETs based on the Potts model for solving complex combinatorial optimization problems.
\newblock \emph{Advanced Materials}, 38(16):e2516797, 2026.

\bibitem{Duffee2025pdits}
C.~Duffee, J.~Athas, A.~Grimaldi, D.~Volpe, G.~Finocchio, E.~Wei, and P.~Khalili~Amiri.
\newblock P-dits: Probabilistic d-dimensional bits for extended-variable probabilistic computing.
\newblock \emph{Physical Review Applied}, 24:044077, 2025.

\bibitem{uva_research_computing}
{University of Virginia Research Computing}.
\newblock UVA Research Computing.
\newblock 2026. [Online]. Accessed: Jun. 21, 2026.
\newblock Available: \url{https://rc.virginia.edu/}

\bibitem{Gset_HelmbergRendl_Stanford}
C.~Helmberg and F.~Rendl.
\newblock Gset: Max-cut benchmark instances.
\newblock 2000. Generated with the machine-independent graph generator RUDY by Giovanni Rinaldi; hosted/maintained by Yinyu Ye. [Online].
\newblock Available: \url{https://web.stanford.edu/~yyye/yyye/Gset/}

\bibitem{khan2025new}
N.~Khan and N.~Shukla.
\newblock New best-known Max-Cut solution for the G63 instance in the G-set benchmark.
\newblock \emph{arXiv preprint} arXiv:2510.21105, 2025.

\bibitem{potts1952generalized}
R.~B. Potts.
\newblock Some generalized order-disorder transformations.
\newblock \emph{Mathematical Proceedings of the Cambridge Philosophical Society}, 48(1):106--109, 1952.

\bibitem{wu1982potts}
F.~Y. Wu.
\newblock The Potts model.
\newblock \emph{Reviews of Modern Physics}, 54(1):235--268, 1982.

\end{thebibliography}

\clearpage
\onecolumngrid  

\setcounter{figure}{0}
\setcounter{table}{0}
\setcounter{equation}{0}

\renewcommand{\thefigure}{S\arabic{figure}}
\renewcommand{\thetable}{S\arabic{table}}
\renewcommand{\theequation}{S\arabic{equation}}

\title{Supplemental Material\\[1.5ex]
Leveraging Population Dynamics to Steer Efficient Search in Large-Scale Combinatorial Optimization\\[0.6ex]}

\author{Nikhat~Khan, Nikhil~Shukla\textsuperscript{*}}

\affiliation{%
\textnormal{\textsuperscript}University of Virginia, Charlottesville, VA, USA\\
\textnormal{\textsuperscript{*}Email: ns6pf@virginia.edu};\;
}

\setcounter{section}{1}
\renewcommand{\thesection}{S\arabic{section}}
\maketitle

\section*{Supplementary Material}
\subsection{Hamming Distance Computation to Measure Replica Diversity}
\label{sec:methods_hamming_distance}

To quantify the diversity among replicas during the optimization process, we computed the pairwise Hamming distance between replica configurations at each stage. For an ensemble of $R$ replicas, the total number of distinct replica pairs is
\[
\frac{R(R-1)}{2}.
\]
For each pair of replicas, we measured the Hamming distance $({d}_{\mathrm H})$ as the number of bit positions at which the two configurations differ. This distance was then normalized by the total number of bits, yielding
\[
d_H \in [0,1],
\]
where $d=0$ indicates identical configurations and $d=1$ indicates configurations that differ at every bit.

Since the Max-Cut objective is invariant under a global bit flip, two complementary configurations represent the same cut. Therefore, a distance close to $1$ may correspond to an equivalent solution rather than a genuinely different one. To account for this symmetry, we defined the corrected distance as
\[
d_H = \min(d,1-d).
\]
This corrected distance lies in the interval
\[
d_H \in [0,0.5].
\]
Values close to $0$ indicate that the replicas correspond to the same or nearly identical cuts, whereas larger values indicate that the replicas occupy genuinely distinct regions of the solution space. The distribution of $d_H$ over all replica pairs was tracked over the course of the run to assess whether the ensemble maintained diversity or collapsed into a common basin.

\vspace{1cm}

\subsection{Detailed results of Augmented PAMC implemented on a GPU}
\label{detailed_results}
\subsubsection{Max-Cut}

{\setlength{\tabcolsep}{8pt}
\renewcommand{\arraystretch}{0.95}
\begin{longtable}{c | c | c | c || c | c || c | c}
\caption{Comparison of SBM and PAMC on benchmark graphs. Comparison of the time-to-solution (TTS) of SBM and PAMC on selected G-set benchmark instances, using the solution reported by SBM as target solution quality. For each graph, the table lists the best-known cut value, the SBM solution used as the target, and the corresponding TTS and success probability $P_s$ for both SBM and PAMC. Statistics are computed over 100 independent runs. Graphs shown in parentheses were evaluated using only 10 independent runs. The PAMC speedup is computed as the ratio of the SBM TTS to the PAMC TTS. Instances in which PAMC achieves a lower TTS than SBM for the same target solution quality are highlighted in bold. Entries marked as ``--'' correspond to instances for which PAMC was unable to reach the target solution quality with the specified parameter settings across all runs.}
\label{tab:maxcut_supplement_table1}\\

\hline
\multirow{2}{*}{Graph} & \multirow{2}{*}{Nodes} & \multirow{2}{*}{Best-known} & \multirow{2}{*}{SBM Sol.} & \multicolumn{2}{c||}{SBM \cite{Goto2021highperformance}} & \multicolumn{2}{c}{PAMC} \\
\cline{5-6} \cline{7-8}
& & & & TTS & Ps & TTS & Ps \\
\hline
\endfirsthead

\caption[]{Comparison of SBM and PAMC on benchmark graphs continued.}\\
\hline
\multirow{2}{*}{Graph} & \multirow{2}{*}{Nodes} & \multirow{2}{*}{Best-known} & \multirow{2}{*}{SBM Sol.} & \multicolumn{2}{c||}{SBM} & \multicolumn{2}{c}{PAMC} \\
\cline{5-6} \cline{7-8}
& & & & TTS & Ps & TTS & Ps \\
\hline
\endhead

\hline
\multicolumn{8}{r}{Continued on next page}\\
\endfoot

\hline
\endlastfoot

G1  & 800   & 11624 & 11624 & 0.033  & 0.9870 & 0.54     & 0.99 \\
G2  & 800   & 11620 & 11620 & 0.048  & 0.8200 & 1.29     & 0.87 \\
G3  & 800   & 11622 & 11622 & 0.003  & 0.9960 & 0.62     & 0.99 \\
G4  & 800   & 11646 & 11646 & 0.005  & 0.9830 & 0.95     & 0.92 \\
G5  & 800   & 11631 & 11631 & 0.012  & 0.9720 & 0.49     & 0.99 \\
G6  & 800   & 2178  & 2178  & 72     & 0.9790 & \textbf{1.00}     & 0.95 \\
G7  & 800   & 2006  & 2006  & 0.34   & 0.9740 & 0.54     & 0.99 \\
G8  & 800   & 2005  & 2005  & 0.35   & 0.9540 & 0.62     & 0.99 \\
G9  & 800   & 2054  & 2054  & 1.6    & 0.8670 & \textbf{1.04}     & 0.96 \\
G10 & 800   & 2000  & 2000  & 0.38   & 0.4070 & 4.24     & 0.53 \\
G11 & 800   & 564   & 564   & 0.018  & 0.9800 & 1.92     & 0.99 \\
G12 & 800   & 556   & 556   & 0.24   & 0.9730 & 1.77     & 0.99 \\
G13 & 800   & 582   & 582   & 0.009  & 0.9960 & 2.00     & 0.99 \\
G14 & 800   & 3064  & 3064  & 0.26   & 0.0050 & 155.78   & 0.11 \\
G15 & 800   & 3050  & 3050  & 0.046  & 0.8040 & 8.06     & 0.85 \\
G16 & 800   & 3052  & 3052  & 0.034  & 0.9920 & 2.24     & 0.99 \\
G17 & 800   & 3047  & 3047  & 0.059  & 0.2830 & 35.48    & 0.25 \\
G18 & 800   & 992   & 992   & 0.006  & 0.0740 & 39.05    & 0.18 \\
G19 & 800   & 906   & 906   & 0.007  & 0.9950 & 2.54     & 0.95 \\
G20 & 800   & 941   & 941   & 0.012  & 0.9800 & 1.50     & 0.99 \\
G21 & 800   & 931   & 931   & 0.036  & 0.1360 & 25.98    & 0.28 \\
G22 & 2000  & 13359 & 13359 & 0.43   & 0.9280 & 75.71    & 0.89 \\
G23 & 2000  & 13344 & 13342 & 0.089  & 0.9890 & 213.70  & 0.98   \\
G24 & 2000  & 13337 & 13337 & 0.46   & 0.6480 & 17.29    & 0.99 \\
G25 & 2000  & 13340 & 13340 & 2.3    & 0.3990 & 434.41   & 0.16 \\
G26 & 2000  & 13328 & 13328 & 0.48   & 0.6430 & 377.28   & 0.18 \\
G27 & 2000  & 3341  & 3341  & 0.05   & 0.9710 & 41.18    & 0.80 \\
G28 & 2000  & 3298  & 3298  & 0.087  & 0.9520 & 25.43    & 0.95 \\
G29 & 2000  & 3405  & 3405  & 0.22   & 0.7370 & 90.12    & 0.54 \\
G30 & 2000  & 3413  & 3413  & 0.44   & 0.7380 & 87.74    & 0.99 \\
G31 & 2000  & 3310  & 3310  & 1.2    & 0.1990 & 93.33    & 0.67 \\
G32 & 2000  & 1410  & 1410  & 3.6    & 0.0930 & 190.94   & 0.90 \\
G33 & 2000  & 1382  & 1382  & 58     & 0.0050 & 2119.67  & 0.34 \\
G34 & 2000  & 1384  & 1384  & 2.1    & 0.2310 & 105.04   & 0.95 \\
G35 & 2000  & 7687  & 7686  & 8319   & 0.0005 & \textbf{348.30}   & 0.60 \\
G36 & 2000  & 7680  & 7680  & 62647  & 0.0001 & \textbf{12361.42} & 0.02 \\
G37 & 2000  & 7691  & 7691  & 27343  & 0.0002 & \textbf{11244.34} & 0.02 \\
G38 & 2000  & 7688  & 7688  & 99     & 0.0680 & 120.36   & 0.78 \\
G39 & 2000  & 2408  & 2408  & 56     & 0.1070 & 147.23   & 0.62 \\
G40 & 2000  & 2400  & 2400  & 24     & 0.1540 & 3924.85  & 0.05 \\
G41 & 2000  & 2405  & 2405  & 11     & 0.2820 & 219.44   & 0.45 \\
G42 & 2000  & 2481  & 2480  & 550    & 0.0001 & 11966.40 & 0.01 \\
G43 & 1000  & 6660  & 6660  & 0.006  & 0.9920 & 2.11     & 0.99 \\
G44 & 1000  & 6650  & 6650  & 0.007  & 0.9850 & 2.31     & 0.99 \\
G45 & 1000  & 6654  & 6654  & 0.043  & 0.9850 & 3.73     & 0.94 \\
G46 & 1000  & 6649  & 6649  & 0.016  & 0.9920 & 2.70     & 0.99 \\
G47 & 1000  & 6657  & 6657  & 0.045  & 0.9820 & 7.05     & 0.83 \\
G48 & 3000  & 6000  & 6000  & 0.0008 & 0.9870 & 36.61    & 0.99 \\
G49 & 3000  & 6000  & 6000  & 0.0008 & 0.9950 & 53.43    & 0.96 \\
G50 & 3000  & 5880  & 5880  & 0.0026 & 1.0000 & 48.38    & 0.99 \\
G51 & 1000  & 3848  & 3848  & 12     & 0.0670 & \textbf{7.25}     & 0.99 \\
G52 & 1000  & 3851  & 3851  & 6.9    & 0.2130 & 38.29    & 0.49 \\
G53 & 1000  & 3850  & 3850  & 94     & 0.0430 & \textbf{65.60}    & 0.32 \\
G54 & 1000  & 3852  & 3852  & 2307   & 0.0010 & \textbf{217.82}   & 0.22 \\
G55 & 5000  & 10299 & 10298 & 6874   & 0.0020 & --       & --   \\
G56 & 5000  & 4017  & 4016  & 6887   & 0.0020 & --       & --   \\
G57 & 5000  & 3494  & 3490  & 6569   & 0.0020 & 10861.40 & 0.99 \\
G58 & 5000  & 19294 & 19283 & 21638  & 0.0010 & \textbf{9062.50}  & 0.93 \\
G59 & 5000  & 6088  & 6086  & 22651  & 0.0010 & --       & --   \\
G60 & 7000  & 14190 & 14186 & 21827  & 0.0010 & 25798.32 & 0.51 \\
G61 & 7000  & 5798  & 5796  & 10883  & 0.0020 & --       & --   \\
G62 & 7000  & 4872  & 4862  & 20501  & 0.0010 & 20931.07 & 0.81 \\
G63 & 7000  & 27045 & 27023 & 31279  & 0.0010 & \textbf{5345.05}  & 0.96 \\
G64 & 7000  & 8752  & 8739  & 31448  & 0.0010 & 75175.96 & 0.15 \\
G65 & 8000  & 5562  & 5546  & 5651   & 0.0040 & 9622.08  & 0.99 \\
G66 & 9000  & 6364  & 6342  & 27408  & 0.0010 & \textbf{7099.79}  & 0.99 \\
G67 & 10000 & 6950  & 6922  & 6340   & 0.0050 & 6411.22  & 0.99 \\
G70 & 10000 & 9595  & 9578  & 31599  & 0.0010 & \textbf{12804.60} & 0.99 \\
G72 & 10000 & 7008  & 6982  & 6142   & 0.0050 & 10265.30 & 0.99 \\
(G77) & 14000 & 9940  & 9904  & 46760 & 0.0010 & \textbf{22225}     & 1.00 \\
(G81) & 20000 & 14060 & 13392 & 62194 & 0.0010 & \textbf{37397}     & 1.00  \\

\end{longtable}

\vspace{2cm}

\subsubsection{Max-3-Cut}

\scriptsize
\setlength{\tabcolsep}{5pt}
\renewcommand{\arraystretch}{0.95}

\begin{longtable}{r r r r r r r r}
\caption{Comparative Max-$3$-Cut (M$3$C) results for G-set instances relative to the previously best-known Max-$3$-Cut solution. $\Delta$ = $M3C_{\text{Our-Work}} - M3C_{\text{Best-Known}}$; ``--'' indicates that the solution to the instance was not reported in the work; MOH: Multi-Operator Heuristic; $n$: nodes; $m$: edges. Entries marked with ($^*$) were obtained using the second update strategy, based on cascaded p-bits for Gibbs sampling.}
\label{tab:max3cut-supplement_table2}\\

\toprule
\multirow{2}{*}{Graph} & \multirow{2}{*}{$n$} & \multirow{2}{*}{$m$} & \multirow{2}{*}{MOH \cite{Ma2017MOH}} & \multicolumn{2}{c}{PLS \cite{garvardt2024parameterized}} & \multirow{2}{*}{Our} & \multirow{2}{*}{$\Delta$} \\
\cmidrule(lr){5-6}
 &  &  &  & LS & ILP &  &  \\
\midrule
\endfirsthead

\caption[]{Comparative Max-$3$-Cut results for Gset instances, continued.}\\
\toprule
\multirow{2}{*}{Graph} & \multirow{2}{*}{$n$} & \multirow{2}{*}{$m$} & \multirow{2}{*}{MOH} & \multicolumn{2}{c}{PLS} & \multirow{2}{*}{Our} & \multirow{2}{*}{$\Delta$} \\
\cmidrule(lr){5-6}
 &  &  &  & LS & ILP &  &  \\
\midrule
\endhead

\midrule
\multicolumn{8}{r}{Continued on next page}\\
\endfoot

\bottomrule
\endlastfoot

G1 & 800 & 19176 & 15165 & -- & -- & 15165 & 0 \\
G2 & 800 & 19176 & 15172 & -- & -- & 15172 & 0 \\
G3 & 800 & 19176 & 15173 & -- & -- & 15173 & 0 \\
G4 & 800 & 19176 & 15184 & -- & -- & 15184 & 0 \\
G5 & 800 & 19176 & 15193 & -- & -- & 15193 & 0 \\
G6 & 800 & 19176 & 2632 & -- & -- & 2632 & 0 \\
G7 & 800 & 19176 & 2409 & -- & -- & 2409 & 0 \\
G8 & 800 & 19176 & 2428 & -- & -- & 2428 & 0 \\
G9 & 800 & 19176 & 2478 & -- & -- & 2478 & 0 \\
G10 & 800 & 19176 & 2407 & -- & -- & 2407 & 0 \\
G11 & 800 & 1600 & 669 & -- & 671 & 671 & 0 \\
G12 & 800 & 1600 & 660 & 661 & 663 & 663 & 0 \\
G13 & 800 & 1600 & 686 & 687 & 688 & 688 & 0 \\
G14 & 800 & 4694 & 4012 & -- & -- & 4012 & 0 \\
G15 & 800 & 4661 & 3984 & 3985 & 3985 & 3983 & -2 \\
G16 & 800 & 4672 & 3991 & -- & -- & \textbf{3992} & 1 \\
G17 & 800 & 4667 & 3983 & -- & -- & \textbf{3986} & 3 \\
G18 & 800 & 4694 & 1207 & -- & -- & 1207 & 0 \\
G19 & 800 & 4661 & 1081 & -- & -- & 1081 & 0 \\
G20 & 800 & 4672 & 1122 & -- & -- & 1122 & 0 \\
G21 & 800 & 4667 & 1109 & -- & -- & \textbf{1110} & 1 \\
G22 & 2000 & 19990 & 17167 & -- & -- & \textbf{17169} & 2 \\
G23 & 2000 & 19990 & 17168 & -- & -- & \textbf{17172} & 4 \\
G24 & 2000 & 19990 & 17162 & 17163 & -- & 17162 & -1 \\
G25 & 2000 & 19990 & 17163 & 17164 & -- & \textbf{17167} & 3 \\
G26 & 2000 & 19990 & 17154 & 17155 & -- & \textbf{17157} & 2 \\
G27 & 2000 & 19990 & 4020 & 4021 & -- & 4021 & 0 \\
G28 & 2000 & 19990 & 3973 & 3975 & -- & \textbf{3978} & 3 \\
G29 & 2000 & 19990 & 4106 & -- & -- & \textbf{4110} & 4 \\
G30 & 2000 & 19990 & 4119 & -- & -- & \textbf{4121} & 2 \\
G31 & 2000 & 19990 & 4003 & 4005 & -- & \textbf{4008} & 3 \\
G32 & 2000 & 4000 & 1653 & 1658 & 1666 & 1664 & -2 \\
G33 & 2000 & 4000 & 1625 & 1628 & 1636 & 1635 & -1 \\
G34 & 2000 & 4000 & 1607 & 1609 & 1616 & 1615 & -1 \\
G35 & 2000 & 11778 & 10046 & 10048 & -- & \textbf{10058} & 10 \\
G36 & 2000 & 11766 & 10039 & -- & -- & \textbf{10047} & 8 \\
G37 & 2000 & 11785 & 10052 & 10053 & 10053 & \textbf{10058} & 5 \\
G38 & 2000 & 11779 & 10040 & -- & -- & \textbf{10054} & 14 \\
G39 & 2000 & 11778 & 2903 & -- & -- & \textbf{2910} & 7 \\
G40 & 2000 & 11766 & 2870 & 2871 & -- & \textbf{2878} & 7 \\
G41 & 2000 & 11785 & 2887 & 2888 & -- & \textbf{2889} & 1 \\
G42 & 2000 & 11779 & 2980 & -- & -- & \textbf{2983} & 3 \\
G43 & 1000 & 9990 & 8573 & -- & -- & 8573 & 0 \\
G44 & 1000 & 9990 & 8571 & -- & -- & 8571 & 0 \\
G45 & 1000 & 9990 & 8566 & -- & -- & 8566 & 0 \\
G46 & 1000 & 9990 & 8568 & -- & -- & 8568 & 0 \\
G47 & 1000 & 9990 & 8572 & -- & -- & 8572 & 0 \\
G48 & 3000 & 6000 & 6000 & -- & -- & 6000* & 0 \\
G49 & 3000 & 6000 & 6000 & -- & -- & 6000* & 0 \\
G50 & 3000 & 6000 & 6000 & -- & -- & 6000* & 0 \\
G51 & 1000 & 5909 & 5037 & -- & -- & \textbf{5039} & 2 \\
G52 & 1000 & 5916 & 5040 & -- & -- & \textbf{5043} & 3 \\
G53 & 1000 & 5914 & 5039 & -- & -- & \textbf{5042} & 3 \\
G54 & 1000 & 5916 & 5036 & -- & -- & \textbf{5040} & 4 \\
G55 & 5000 & 12498 & 12429 & 12429 & 12432 & 12432 & 0 \\
G56 & 5000 & 12498 & 4752 & 4757 & -- & \textbf{4763} & 6 \\
G57 & 5000 & 10000 & 4083 & 4092 & 4103 & \textbf{4114} & 11 \\
G58 & 5000 & 29570 & 25195 & -- & -- & \textbf{25216} & 21 \\
G59 & 5000 & 29570 & 7262 & 7276 & -- & \textbf{7309} & 33 \\
G60 & 7000 & 17148 & 17076 & -- & -- & \textbf{17082} & 6 \\
G61 & 7000 & 17148 & 6853 & 6861 & -- & \textbf{6872} & 11 \\
G62 & 7000 & 14000 & 5685 & 5710 & 5706 & \textbf{5731} & 21 \\
G63 & 7000 & 41459 & 35322 & 35318 & -- & \textbf{35355} & 33 \\
G64 & 7000 & 41459 & 10443 & 10437 & -- & \textbf{10508} & 65 \\
G65 & 8000 & 16000 & 6490 & 6512 & 6535 & \textbf{6544} & 9 \\
G66 & 9000 & 18000 & 7416 & 7442 & 7443 & \textbf{7475} & 32 \\
G67 & 10000 & 20000 & 8086 & 8116 & 8141 & \textbf{8152} & 11 \\
G70 & 10000 & 9999 & 9999 & -- & -- & 9999* & 0 \\
G72 & 10000 & 20000 & 8192 & 8244 & 8244 & \textbf{8259} & 15 \\

\end{longtable}
\normalsize

\vspace{1cm}

\subsection{Derivation of a three-state cascaded p-bit sampler for the Potts model}
\label{derivation_potts}

This Supplementary Note derives the cascaded binary p-bit sampler used to implement the exact (K=3) Potts Gibbs update. The goal is to show that a three-state Potts variable can be sampled exactly through two sequential binary stochastic decisions, enabling a binary-compatible realization of the Max-3-Cut update rule.

Consider a three-state Potts variable at node $i$,
\begin{equation}
    \sigma_i \in \{\alpha,\beta,\gamma\}.
\end{equation}
That is, node $i$ may occupy one of the three possible states
$\alpha$, $\beta$, or $\gamma$. The corresponding Gibbs conditional probabilities are
\begin{align}
    P(\sigma_i^+=\alpha)
    &=
    \frac{\omega_i^\alpha}
    {\omega_i^\alpha+\omega_i^\beta+\omega_i^\gamma},
    \label{eq:gibbs_alpha_three_state}
    \\
    P(\sigma_i^+=\beta)
    &=
    \frac{\omega_i^\beta}
    {\omega_i^\alpha+\omega_i^\beta+\omega_i^\gamma},
    \label{eq:gibbs_beta_three_state}
    \\
    P(\sigma_i^+=\gamma)
    &=
    \frac{\omega_i^\gamma}
    {\omega_i^\alpha+\omega_i^\beta+\omega_i^\gamma}.
    \label{eq:gibbs_gamma_three_state}
\end{align}
Here,
\begin{equation}
    \omega_i^\alpha = e^{-\rho E_i^\alpha},
    \qquad
    \omega_i^\beta = e^{-\rho E_i^\beta},
    \qquad
    \omega_i^\gamma = e^{-\rho E_i^\gamma},
    \label{eq:three_state_boltzmann_weights}
\end{equation}
are the Boltzmann weights associated with assigning node $i$ to
states $\alpha$, $\beta$, and $\gamma$, respectively. The parameter
$\rho$ denotes the inverse temperature.

The goal is to sample from the distribution in
Eqs.~\eqref{eq:gibbs_alpha_three_state}--\eqref{eq:gibbs_gamma_three_state}
using a cascade of binary p-bits. The first binary decision determines
whether state $\alpha$ is selected. If $\alpha$ is rejected, a second
binary decision determines whether state $\beta$ or $\gamma$ is selected.

Starting with the probability of choosing state $\alpha$, we have
\begin{equation}
    P(\sigma_i^+=\alpha)
    =
    \frac{\omega_i^\alpha}
    {\omega_i^\alpha+\omega_i^\beta+\omega_i^\gamma}.
\end{equation}
Dividing the numerator and denominator by $\omega_i^\alpha$ gives
\begin{equation}
    P(\sigma_i^+=\alpha)
    =
    \frac{1}
    {1+\dfrac{\omega_i^\beta+\omega_i^\gamma}{\omega_i^\alpha}}.
\end{equation}
Equivalently,
\begin{equation}
    P(\sigma_i^+=\alpha)
    =
    \frac{1}
    {1+
    \exp\left[
        \ln(\omega_i^\beta+\omega_i^\gamma)
        -
        \ln \omega_i^\alpha
    \right]}.
    \label{eq:alpha_logistic_three_state}
\end{equation}
Using the identity
\begin{equation}
    \frac{1}{1+e^{-x}}
    =
    \frac{1}{2}
    \left[
        1+\tanh\left(\frac{x}{2}\right)
    \right],
    \label{eq:logistic_tanh_identity_three_state}
\end{equation}
define the first p-bit input as
\begin{equation}
    \Delta_1
    =
    \frac{1}{2}
    \left[
        \ln \omega_i^\alpha
        -
        \ln(\omega_i^\beta+\omega_i^\gamma)
    \right].
    \label{eq:delta1_three_state}
\end{equation}
Then
\begin{equation}
    P(\sigma_i^+=\alpha)
    =
    \frac{1}{2}
    \left[
        1+\tanh(\Delta_1)
    \right].
    \label{eq:alpha_tanh_three_state}
\end{equation}
Let $\eta_i^{(1)}$ be uniformly distributed on $[-1,1]$. The first
binary p-bit is defined by
\begin{equation}
    b_1
    =
    \frac{1+
    \operatorname{sgn}
    \left[
        \tanh(\Delta_1)-\eta_i^{(1)}
    \right]}
    {2},
    \qquad
    \eta_i^{(1)}\sim \mathrm{Uniform}[-1,1].
    \label{eq:b1_three_state}
\end{equation}
Thus, $b_1=1$ corresponds to selecting state $\alpha$.
If state $\alpha$ is rejected, then the sampler chooses between states
$\beta$ and $\gamma$. The conditional probability of choosing $\beta$,
given that $\alpha$ was not selected, is
\begin{equation}
    P(\sigma_i^+=\beta \mid \sigma_i^+\neq \alpha)
    =
    \frac{\omega_i^\beta}
    {\omega_i^\beta+\omega_i^\gamma}.
    \label{eq:beta_conditional_three_state}
\end{equation}
Define the second p-bit input as
\begin{equation}
    \Delta_2
    =
    \frac{1}{2}
    \left[
        \ln \omega_i^\beta
        -
        \ln \omega_i^\gamma
    \right].
    \label{eq:delta2_three_state}
\end{equation}
Then
\begin{equation}
    P(\sigma_i^+=\beta \mid \sigma_i^+\neq \alpha)
    =
    \frac{1}{2}
    \left[
        1+\tanh(\Delta_2)
    \right].
\end{equation}
Let $\eta_i^{(2)}$ be another independent random variable uniformly
distributed on $[-1,1]$. The second binary p-bit is
\begin{equation}
    b_2
    =
    \frac{1+
    \operatorname{sgn}
    \left[
        \tanh(\Delta_2)-\eta_i^{(2)}
    \right]}
    {2},
    \qquad
    \eta_i^{(2)}\sim \mathrm{Uniform}[-1,1].
    \label{eq:b2_three_state}
\end{equation}
The resulting cascaded three-state update is
\begin{equation}
    \sigma_i^+
    =
    \alpha b_1
    +
    (1-b_1)
    \left[
        \beta b_2
        +
        \gamma(1-b_2)
    \right].
    \label{eq:three_state_cascaded_update}
\end{equation}
Equivalently,
\begin{equation}
    \sigma_i^+
    =
    \begin{cases}
        \alpha, & b_1=1,\\
        \beta,  & b_1=0,\ b_2=1,\\
        \gamma, & b_1=0,\ b_2=0.
    \end{cases}
    \label{eq:three_state_branching_update}
\end{equation}
The output probabilities are
\begin{align}
    P(\sigma_i^+=\alpha)
    &= P(b_1=1)
    =
    \frac{\omega_i^\alpha}
    {\omega_i^\alpha+\omega_i^\beta+\omega_i^\gamma},
    \\
    P(\sigma_i^+=\beta)
    &= P(b_1=0)P(b_2=1) \nonumber\\
    &=
    \left(
        \frac{\omega_i^\beta+\omega_i^\gamma}
        {\omega_i^\alpha+\omega_i^\beta+\omega_i^\gamma}
    \right)
    \left(
        \frac{\omega_i^\beta}
        {\omega_i^\beta+\omega_i^\gamma}
    \right) \nonumber\\
    &=
    \frac{\omega_i^\beta}
    {\omega_i^\alpha+\omega_i^\beta+\omega_i^\gamma},
    \\
    P(\sigma_i^+=\gamma)
    &= P(b_1=0)P(b_2=0) \nonumber\\
    &=
    \left(
        \frac{\omega_i^\beta+\omega_i^\gamma}
        {\omega_i^\alpha+\omega_i^\beta+\omega_i^\gamma}
    \right)
    \left(
        \frac{\omega_i^\gamma}
        {\omega_i^\beta+\omega_i^\gamma}
    \right) \nonumber\\
    &=
    \frac{\omega_i^\gamma}
    {\omega_i^\alpha+\omega_i^\beta+\omega_i^\gamma}.
\end{align}
Thus, the cascaded p-bit update exactly samples from the desired
three-state Gibbs conditional distribution.
Using $\omega_i^a=e^{-\rho E_i^a}$, we have
\begin{equation}
    \ln \omega_i^a = -\rho E_i^a.
\end{equation}
Therefore, the second p-bit input can be written as
\begin{equation}
    \Delta_2
    =
    \frac{1}{2}
    \left[
        \ln\omega_i^\beta-\ln\omega_i^\gamma
    \right]
    =
    -\frac{\rho}{2}
    \left[
        E_i^\beta-E_i^\gamma
    \right].
\end{equation}
Thus, $\Delta_2$ compares the relative energetic preference for
$\beta$ versus $\gamma$.

Similarly,
\begin{equation}
    \Delta_1
    =
    \frac{1}{2}
    \left[
        \ln \omega_i^\alpha
        -
        \ln(\omega_i^\beta+\omega_i^\gamma)
    \right].
\end{equation}
Thus, $\Delta_1$ compares the weight of state $\alpha$ against the
combined weight of the two remaining alternatives, $\beta$ and $\gamma$.

\end{document}